\documentclass[Conference,a4paper]{IEEEtran}
\IEEEoverridecommandlockouts
\usepackage{color}
\usepackage{graphicx}
\usepackage{epstopdf}
\usepackage{amsmath}
\usepackage{amssymb}
\usepackage{algorithm}
\usepackage{algorithmic}
\usepackage{amsmath}
\usepackage{multirow}
\usepackage{booktabs}
\usepackage{array}
\usepackage{amsthm}
\usepackage{stfloats}
\usepackage{caption}
\usepackage{subfigure}
\usepackage{bm}
\usepackage{booktabs}
\usepackage{setspace}
\usepackage{diagbox}
\usepackage{enumerate}
\usepackage{ulem}
\usepackage{url}
\usepackage{circledsteps}

\usepackage{hyperref}
\allowdisplaybreaks[4]

{\bgroup
 \addtolength\abovedisplayshortskip{#1}
 \addtolength\abovedisplayskip{#1}
 \addtolength\belowdisplayshortskip{#1}
 \addtolength\belowdisplayskip{#1}
 }
{\egroup\ignorespacesafterend}

\newtheorem{theorem}{Theorem}

\newcommand{\be}{\begin{equation}}
\newcommand{\ee}{\end{equation}}
\newcommand{\bea}{\begin{eqnarray}}
\newcommand{\eea}{\end{eqnarray}}
\newcommand{\ba}{\begin{array}}
\newcommand{\ea}{\end{array}}

\title{
Beamforming for
Transmissive RIS Transceiver Enabled Simultaneous Wireless Information and Power Transfer Systems
}
\author{\IEEEauthorblockN{Yuan Guo, Wen Chen, Yanze Zhu, Zhendong Li, Qiong Wu, and Kunlun Wang
}
\thanks{
Y. Guo, W. Chen and
Y. Zhu are with Department of Electronic Engineering, Shanghai Jiao Tong University, Shanghai, China, 
email:
yuanguo26@sjtu.edu.cn,
wenchen@sjtu.edu.cn,
yanzezhu@sjtu.edu.cn.}
\thanks{Z. Li is with the School of Information and Communication Engineering, Xi'an Jiaotong University, Xi'an, China,
email:
lizhendong@xjtu.edu.cn.}
\thanks{
Q. Wu is with the School of Internet of Things Engineering, 
Jiangnan University, Wuxi, China, 
emali: qiongwu@jiangnan.edu.cn.
}
\thanks{
K. Wang is with 
the School of Communication and Electronic Engineering, 
East China Normal University, Shanghai, China,
email: klwang@cee.ecnu.edu.cn.
}

}

\begin{document}
\maketitle
\pagestyle{empty}
\thispagestyle{empty}

\begin{abstract}
This paper investigates a novel transmissive reconfigurable intelligent surface (TRIS) transceiver-empowered
simultaneous wireless information and power transfer (SWIPT) system 
with multiple information decoding (ID) and energy harvesting (EH) users.
Under the considered system model, 
we formulate an optimization problem 
that maximizes the sum-rate of all ID users 
via the design of the TRIS transceiver's active beamforming. 
The design is constrained by per-antenna power limits at the TRIS transceiver 
and by the minimum harvested energy demand of all EH users.
Due to the non-convexity of the objective function and the energy harvesting constraint,
the sum-rate problem is difficult to tackle.
To solve this challenging optimization problem,
by leveraging the weighted minimum mean squared error (WMMSE) framework
and the majorization-minimization (MM) method,
we propose a second-order cone programming (SOCP)-based algorithm.
Per-element power constraints introduce a large number of constraints, 
making the problem considerably more difficult.
By applying the alternating direction method of multipliers (ADMM) method,
we successfully develop an analytical, computationally efficient, 
and highly parallelizable algorithm to address this challenge.
Numerical results are provided to validate the convergence and effectiveness of the proposed algorithms. 
Furthermore, 
the low-complexity algorithm significantly reduces computational complexity without performance degradation. 

\end{abstract}

\begin{IEEEkeywords}
Simultaneous wireless information and power transfer (SWIPT),
transmissive reconfigurable intelligent surface (TRIS) transceiver,
per-element power constraint,
analytic-based solution.
\end{IEEEkeywords}

\maketitle
\section{Introduction}

With the explosive growth in the number of Internet-of-Things (IoT) devices 
and the continuously increasing demand for energy harvesting (EH),
simultaneous wireless information and power transfer (SWIPT) 
\cite{ref_SWIPT_1}$-$\cite{ref_SWIPT_2},
which enables the dual use of radio frequency (RF) signals for both information delivery and energy charging,
has emerged as a highly promising technique for future wireless networks. 
Specifically,
as an operational mode for SWIPT systems, 
a power-supplied base station (BS) transmits RF signals to two groups of devices.
One group of devices,
known as information decoding (ID) users,
 can decode the information form the received signals,
 while the other group of devices, 
referred to as EH users, 
 will harvest energy from those signals.
Besides,
the EH users generally require a substantially higher received power level compared to the ID users.

Recently,
reconfigurable intelligent surface (RIS) has rsien a cost-efficient solution 
for improving wireless network performance by enhancing passive beamforming gain \cite{ref_RIS_1}$-$\cite{ref_RIS_2}.
In general, 
an RIS is a planar array composed of a vast number of low-cost and passive tunable elements, 
which are made of meta-materials
and each capable of dynamically adjusting the phase shifts of the electromagnetic (EM) waveforms incident upon it.
The phase shifts of all reflective elements can be jointly adjusted 
so that reflected signals combine constructively to boost the desired BS signal 
or destructively to suppress interference.
Its flexible phase-shifting control can achieve fine-grained passive beamforming gain
and  
thereby reconfiguring wireless propagation channels to achieve significant performance gain in wireless networks.

In light of the aforementioned merits of RIS architecture,
a growing body of literature has studied employing the RIS to into various wireless systems for enhancing performance,
e.g., \cite{ref_RIS_app_1}$-$\cite{ref_RIS_SWIPT_7}. 
For instance, 
the authors of \cite{ref_RIS_app_1} investigated an RIS-aided downlink multi-group multi-cast system,
formulating a sum-rate maximization problem of all the multi-casting groups,
and developed two efficient algorithms to handle the non-convex problem.
The paper \cite{ref_RIS_app_2} 
considered the sum-mean-square-error (sum-MSE) minimization problem in the millimeter wave (mmWave)
multi-user multiple-input multiple-output (MU-MIMO) system assisted by the RIS
and validated that a large-scale Kronecker-structured hybrid array 
can effectively substitute the RIS in supporting mmWave communications.
The work \cite{ref_RIS_app_3} studied a novel double-faced active (DFA)-RIS architecture that 
mitigates severe ``double-fading'' loss and provides full-space coverage, 
and employed the DFA-RIS to enhance physical-layer security.
A deep learning-based end-to-end (E2E) optimization framework for improving the system spectral efficiency 
in the near-field wideband RIS assisted MIMO system was researched in \cite{ref_RIS_app_4}.
In \cite{ref_RIS_app_5}, 
the authors adopted an RIS for a full-duplex (FD) integrated sensing and communication (ISAC) system 
to improve radar detection probability by suppressing self-interference, 
and developed a highly efficient and low-complexity algorithm with closed-form updates for all variables.
The literature \cite{ref_RIS_app_6} presented a novel intelligent omni surface (IOS)-aided ISAC system
for a multi-user and multi-target scenario
that seeks to maximize the minimum sensing signal-to-interference-plus-noise-ratio (SINR) 
while meeting required communication performance levels.
The paper \cite{ref_RIS_app_7}  innovatively adopted communication data to enhance 
channel state information (CSI) recovery efficiency
in an uplink MU-MIMO network aided by an active RIS.

Meanwhile, 
RIS-aided SWIPT systems 
have also attracted considerable attention in the literature, e.g., \cite{ref_RIS_SWIPT_1}$-$\cite{ref_RIS_SWIPT_7}.
For example,
the authors of \cite{ref_RIS_SWIPT_1} 
deployed an RIS to assist a SWIPT system
maximizing the total harvested power while ensuring the individual SINR requirement.
The paper \cite{ref_RIS_SWIPT_2} 
considered a SWIPT system assisted by multiple RISs
and formulated the problem of minimizing transmit power 
subject to the quality-of-service (QoS) constraints for all information users 
and energy-harvesting constraints for energy users.
The work \cite{ref_RIS_SWIPT_3} 
proposed using  the RIS to enhance the weighted sum-rate of information users 
while guaranteeing the energy harvesting requirement of the energy users in a SWIPT system.
The literature \cite{ref_RIS_SWIPT_4}
introduced a new network architecture for RIS-assisted SWIPT with non-orthogonal multiple access (NOMA), 
where the RIS is leveraged to boost both NOMA throughput and the wireless power transfer (WPT) efficiency.
In \cite{ref_RIS_SWIPT_5}, 
an RIS-assisted secure SWIPT network comprising multiple information users and energy users was studied, 
where the objective is to maximize the weighted sum of transferred power under secrecy-rate constraints.
The authors of  \cite{ref_RIS_SWIPT_6} 
designed a simultaneously transmitting and reflecting  (STAR)-RIS aided SWIPT system
and evaluate three STAR-RIS protocols 
(i.e., 
the energy splitting
(ES), the mode switching (MS), and the time switching (TS) protocols) 
aimed at enhancing the weighted sum harvested power.
The paper \cite{ref_RIS_SWIPT_7} 
studied the near- and hybrid-field channel models for RIS-reflected in 
an interference-limited SWIPT system,
employing independent power splitting model, 
assisted by dual RISs.

Beyond the conventional use of RIS as an auxiliary element in wireless networks,
the recent work \cite{ref_TRIS_1} proposed a novel \textit{TRIS transceiver} architecture
that yields higher system performance with lower power consumption.
The TRIS transceiver integrates a passive TRIS and a single horn antenna feed.
Therefore,
compared with the conventional multi-antenna systems that rely on active components,
the TRIS transceiver obviates the requirement for numerous RF chains and sophisticated signal-processing modules.
Besides,
compared with the reflective RIS transmitter proposed in \cite{ref_TRIS_2}$-$\cite{ref_TRIS_3}, 
the TRIS transceiver technique effectively overcomes two main challenges:
1) feed-source blockage: 
When the horn antenna and the user are on the same side of a reflective RIS, 
the incident EM wave can be occluded by the feed source.
By placing the horn antenna and the user on opposite sides of the surface, 
the TRIS transceiver avoids this blockage;
2)
echo interference:
Since both incident and reflected waves coexist on the same side of a reflective RIS, 
reflective-type transceiver is vulnerable to echo interference. 
The TRIS transceiver mitigates this problem by spatially separating incident and transmitted waves on opposite sides of the RIS. 
Consequently, 
the TRIS transceiver constitutes a promising and cost-effective approach to sustaining capacity growth.

Nowadays,
great attention has been paid to the integration of TRIS transceiver into wireless networks 
from various perspectives to improve overall system performance, 
e.g., \cite{ref_TRIS_app_1}$-$\cite{ref_TRIS_app_10}.
The authors of \cite{ref_TRIS_app_1} studied a linear complexity algorithm to solve the max-min SINR problem 
in a TRIS transceiver-aided multi-stream downlink communication system employing time-modulated array (TMA).
Under the case of imperfect channel state information (CSI),
the paper \cite{ref_TRIS_app_2} 
investigated the sum-rate maximization problem for a TRIS transceiver-enabled SWIPT system 
with a nonlinear energy-harvesting model.
The work \cite{ref_TRIS_app_3} considered a novel multi-tier computing system assisted by the TRIS transceiver,
aimed at improving computing capabilities, 
decreasing computation latency, 
and reducing BS deployment costs.
The literature \cite{ref_TRIS_app_4} proposed a hybrid active-passive-type TRIS transceiver design 
that allows individual RIS elements to switch dynamically between active and passive modes,
and validated this architecture markedly enhances system energy efficiency (EE).
The authors in \cite{ref_TRIS_app_5} jointly employed TRIS transceiver and RIS to 
enhance the weighted sum secrecy rate in a secure communication system under imperfect reflection CSI.
In \cite{ref_TRIS_app_6},
based on the rate-splitting multiple access (RSMA) technology,
a time-division sensing communication framework was proposed for a TRIS transceiver-enabled secure ISAC system.
The authors of \cite{ref_TRIS_app_7} designed a distributed cooperative ISAC system 
empowered by the TRIS transceiver to extend service coverage and employed the RSMA technology at the BS to enable cooperation.
The paper \cite{ref_TRIS_app_8} 
studied the sum-rate maximization 
for multi-cluster Low Earth
Orbit (LEO) NOMA satellite systems by employing the TRIS transceiver architecture.
The work \cite{ref_TRIS_app_9} 
adopted the TRIS transceiver-enabled spatial modulation (SM) MIMO system
with the maximum likelihood detection at the receiver,
and derived closed-form upper-bound expressions for the average bit-error probability (ABEP).
The authors in \cite{ref_TRIS_app_10} proposed three algorithms to solve the max-min sum-rate problem in a TRIS transceiver-enabled 
multi-group multi-cast downlink communication system.

While the existing works \cite{ref_RIS_SWIPT_1}$-$\cite{ref_RIS_SWIPT_7}  
have mainly investigated reflective-type RIS-aided SWIPT networks,
the integration of TRIS transceiver into SWIPT networks is still in its infancy.
Note that the work \cite{ref_TRIS_app_2} considered the TRIS transceiver-enabled SWIPT system with 
the power splitting  architecture,
in which the receiver divides the received RF signal into two streams according to a PS ratio.
Importantly, 
a separated receiver architecture of the SWIPT system enabled by TRIS transceiver,
where energy harvesting and information decoding are performed by separate receivers,
has not been considered in existing works.

Motivated by the above observations, 
we aim to enhance the performance of SWIPT systems by leveraging TRIS transceiver architecture.
Towards this end, 
this paper considers a TRIS transceiver-enabled SWIPT network,
in which the TRIS transceiver simultaneously serves multiple ID and EH users. 
Specifically,
the contributions of this paper are elaborated as follows:
\begin{itemize}
\item
This paper studies beamforming strategies for a SWIPT system with multiple ID and EH users, 
enabled by the TRIS transceiver architecture,
with the objective of potential performance improvements.
Specifically,
we formulate the sum-rate maximization problem for optimizing the TRIS transceiver beamforming
subject to the minimum harvested energy of all EH users 
and the individual maximum transmit power limits of each TRIS transceiver units.
Note that imposing per-element power constraints substantially 
increases the number of constraints and makes the optimization problem highly challenging.
To the best of our knowledge, 
both the system model and the per-element power constraints considered in this work 
have not been studied in the existing literature.

\item
Due to the non-convexity of both the objective function and the energy-harvesting constraint,
the sum-rate maximization optimization problem is difficult to solve.
To make this problem more tractable,
by leveraging the  weighted minimum mean squared error (WMMSE) framework \cite{ref_WMMSE},
we firstly reformulate the objective function into an equivalently form.
And then,
by integrating the  majorization-minimization (MM) method \cite{ref_MM},
the non-convex optimization problem can be transformed into a second-order cone programming (SOCP) problem.

\item
To reduce computational complexity,
we propose an efficient algorithm that does not resort to any numerical solver.
Specifically,
by applying the alternating direction method of multipliers (ADMM) methodology \cite{ref_ADMM}
and an analysis of the optimality conditions,
we derive closed-form updates for every variable block, and these updates can be computed in parallel.

\item
Last but not least,
extensive numerical results are provided to demonstrate 
the effectiveness and efficiency of the proposed algorithms under various system setups.
The results demonstrate that the ADMM-based algorithm 
outperforms the SOCP-based approach in terms of computational complexity 
without causing performance loss in sum-rate.

\end{itemize}

The remainder of the paper is organized as follows. 
In Section II,
we introduce the system model of the TRIS transceiver-enabled SWIPT system and formulate the sum-rate maximization problem.
Section III proposes the SOCP-based algorithm for the formulated sum-rate optimization problem.
In Section IV,
a low-complexity method based on the ADMM framework will be developed.
Section V presents the simulation results for assessing the effectiveness of 
the proposed algorithms. 
Finally, Section VI presents the conclusions of the paper.

Lower-case and boldface capital letters are respectively represented as  vectors and matrices;
Lowercase, bold lowercase, and bold uppercase letters are respectively represented as
scalars, vectors, and matrices;
$\mathbb{C}^{N \times M}$ represents the set of $N \times M$ complex-valued matrices;
$\mathbf{X}^{\ast}$,
$\mathbf{X}^{T}$,
and
$\mathbf{X}^{H}$
denote the conjugate, transpose, and
conjugate transpose of matrix $\mathbf{X}$, respectively;

\section{System Model and Problem Formulation}
\subsection{System Model}

\begin{figure}[t]
	\centering
	\includegraphics[width=.40\textwidth]{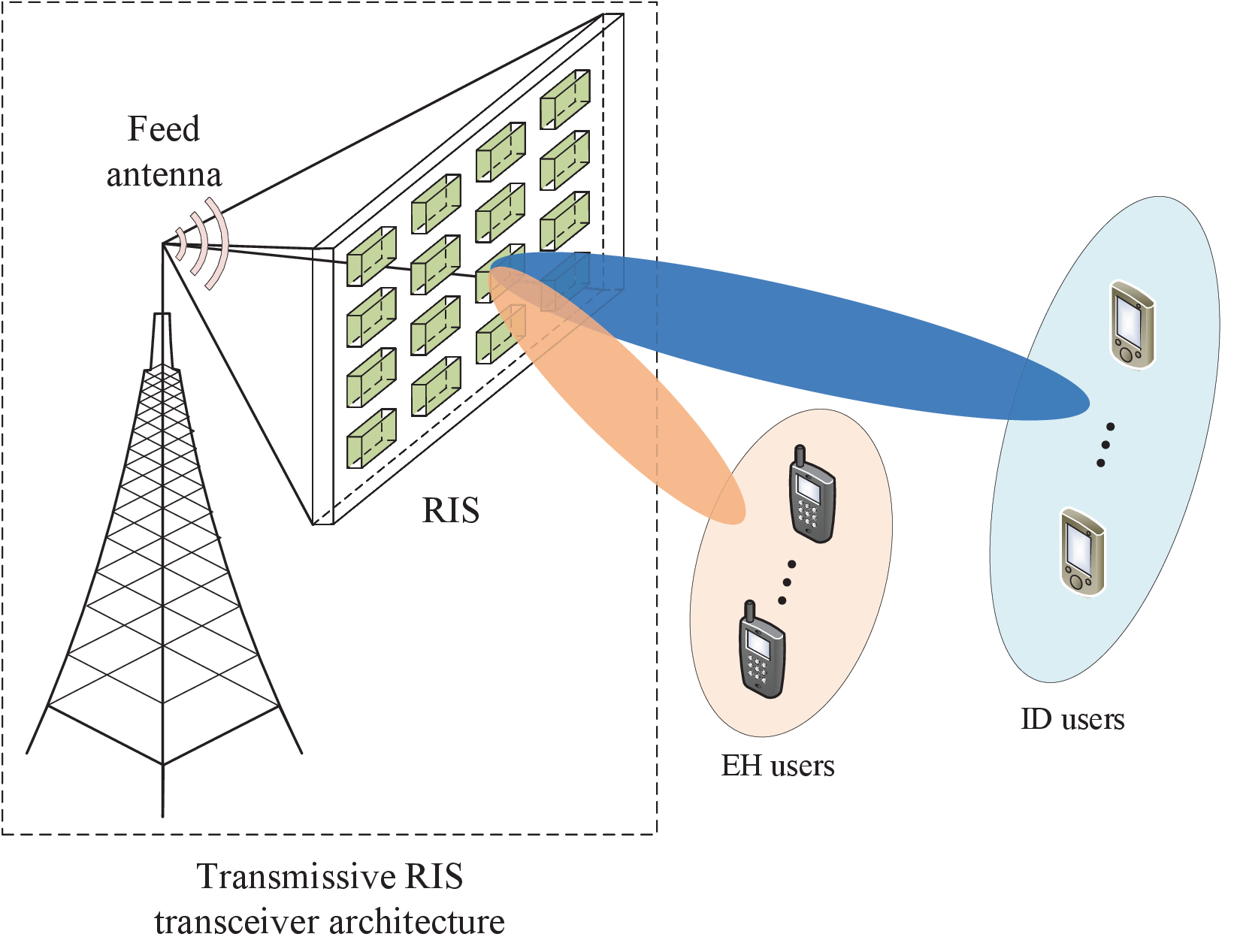}
	\caption{An illustration of the TRIS transceiver-enabled SWIPT system.}
	\label{fig.1}
\end{figure}

We consider a TRIS transceiver-enabled SWIPT system as shown in Fig. \ref{fig.1},
where a TRIS transceiver equipped with $N$ antennas simultaneously serves
two groups of single-antenna users,
i.e.,
 $K$ single-antenna ID users 
and $G$ single-antenna EH users.
Specifically,
for simplicity,
the sets of ID users, EH users, and TRIS transceiver units are respectively represented as
$\mathcal{K} \triangleq \{1,\cdots,K\}$,
$\mathcal{G} \triangleq \{1,\cdots,G\}$,
and
$\mathcal{N} \triangleq \{1,\cdots,N\}$.

Let 
$x_{I,k} \in \mathbb{C}$ 
and
$x_{E,g} \in \mathbb{C}$
denote the modulated information-bearing and energy-carrying signals,
respectively,
which both can be signals of any modulation order.
The information signals are assumed to be independent and identically distributed (i.i.d.) 
CSCG random variables with zero mean and unit variance,
i.e.,
$x_{I,k} \sim  \mathcal{CN}(0,1)$.
Besides,
energy signals are pseudorandom signals that do not contain any information.
Therefore, 
we assumed that energy signals are independently 
produced from an arbitrary distribution such that $\mathbb{E}\{ \vert x_{E,g}  \vert^2 \} = 1$.
Without loss of generality, 
assume each ID user is assigned an individual information beam, 
and each EH user is assigned an individual energy beam.
Moreover,
the vectors 
$\mathbf{f}_{I,k}\in \mathbb{C}^{N\times 1}$
and 
$\mathbf{f}_{E,g}\in \mathbb{C}^{N\times 1}$
denote the beamforming for the $k$-th ID user and the $g$-th EH user, 
respectively.

Then,
the transmitted signal by the TRIS architecture is given by 
\begin{align}
\mathbf{x} = \sum_{k=1}^{K} \mathbf{f}_{I,k}x_{I,k}
+
\sum_{g=1}^{G} \mathbf{f}_{E,g}x_{E,g}.
\end{align}

Furthermore,
any element of this vector $\mathbf{x}$ can be given as
\begin{align}
\mathbf{x}[n]
&=
a_{n,1}b_{1}e^{j(\varphi_{n,1}+ \varpi_{1} )} 
+ \cdots +
a_{n,K}b_{K}e^{j(\varphi_{n,K}+ \varpi_{K} )} \\
&
+a_{n,K+1}b_{K+1}e^{j(\varphi_{n,K+1}+ \varpi_{K+1} )}\nonumber\\
&+ \cdots +
a_{n,K+G}b_{K+G}e^{j(\varphi_{n,K+G}+ \varpi_{K+G} )}=
A_{n}e^{j\phi_n} \nonumber,
\end{align}
where $A_{n}$ and $e^{j\phi_n}$ are the signal amplitude and phase of $\mathbf{x}[n]$, respectively.
According to \cite{ref_TMA}, 
the TMA can generate the $n$-th control signal, 
which corresponds to the $n$-th transmissive unit of the TRIS architecture, 
for the signal vector $\mathbf{x}$.

In the following,
by using TMA-based operation,
we demonstrate a 1-bit RIS example 
that illustrates how the synthesized modulation symbol $\mathbf{x}[n]$, 
which is characterized by the amplitude $A_{n}$ and phase $e^{j\phi_n}$,
is converted into control-signal waveforms corresponding to binary states, i.e., the ``0'' and ``1'' states.
The TRIS element produces phase deviations of 0 for the 0-state and $\pi$ for the 1-state. 
Denote the control-signal symbol duration by $T_c$.
The 0-state control pulse is initiated at $t_o$ where $ 0 \leq t_o \leq T_c$ 
and persists for a duration $ \tau$ with $ 0 \leq \tau \leq T_c$. 
The remaining interval $T_c-\tau$ constitutes the on-period of the 1-state.

To begin with,
$T_c$ is linked to both the switch dynamics of the array and the count of phase levels,
we adopt $T_c=3 \mu\text{s}$ for the 1-bit RIS case. 
Next, 
the time parameter $t_o$ is governed by the array's design and is picked so the 0-state commences at the time 
that yields the intended radiation pattern.
Finally, 
$\tau$ is selected through a trade-off between sidelobe suppression 
and overall radiation-pattern performance.

Based on the above discussion, 
the following two possible situations will arise depending on whether 
 $t_o+\tau$
is less than or greater than $T_c$:

\begin{itemize}
\item[] {CASE-1}:
When $t_o+\tau < T_c$,
the phase shift of the transmission EM wave is given as
\begin{align}
s_n(t)= \label{TMA_EM_1}
\begin{cases}
e^{j0}, & t_{o,n}<t\leq t_{o,n}+\tau_{n},\\
e^{j\pi}, & \text{otherwise}.
\end{cases}
\end{align}

\item[] {CASE-2}: 
When $t_0+\tau > T_c$,
the phase shift of the transmission EM wave is the cyclically shifted version of equation (\ref{TMA_EM_1}), 
which can be given as
\begin{align}
s_n(t)= 
\begin{cases}
e^{j\pi}, & t_{o,n}+\tau_{n} - T_c<t\leq t_{o,n},\\
e^{j0}, & \text{otherwise}.
\end{cases}
\end{align}

\end{itemize}

Furthermore,
by using the Fourier transform for the phase control function of the transmission EM wave, 
we can have
\begin{align}
&S_n(j2\pi f)
=
\frac{1}{T_c}\int_{0}^{T_c}s_n(t)e^{-j2\pi f t}\,dt\\
&=
\begin{cases}
\alpha(f, t_{o,n}, \tau_n) + \frac{j(1-e^{-j2\pi f T_c})}{2\pi f T_c},t_{o,n}+\tau_n \leq T_c, \\
\alpha(f, t_{o,n}, \tau_n-T_c) + \frac{j(e^{-j2\pi f T_c}) - 1}{2\pi f T_c},t_{o,n}+\tau_n > T_c,
\end{cases}\nonumber
\end{align}
where the function $\alpha(f, t_{o,n}, \tau_n)$ is formulated as
\begin{align}
\alpha(f, t_{o,n}, \tau_n)
=
\frac{2}{\pi f T_c} \sin(\pi f T_c)e^{-j \pi f (2t_{o,n}+\tau_n)}.
\end{align}

Following the Fourier series expansion used for time modulation, 
the peak magnitude of the $q$-th harmonic component of the phase modulation function 
is given by
\begin{align}
\hat{S}_{n}\bigg(j 2 \pi \frac{q}{T_c}\bigg)
=
\frac{2}{\pi q }\sin\bigg( \frac{\pi q \tau_n}{T_c} \bigg)
e^{-j \pi q( 2t_{o,n}+\tau_n )/T_c}.
\end{align}

It is important to recognize 
that harmonic component can effectively represent 
both the amplitude and the phase shift of the synthesized modulation symbol $\mathbf{x}[n]$.
Since the positive and negative first-harmonic components contain the largest energy level, 
$\mathbf{x}[n]$  is best represented by one of these first harmonics. 
Taking the positive first harmonic as an example, 
the symbol parameters $A_n$ and $\phi_n$ yield 
the associated control-signal timing values $t_{o,n}$ and $\tau_n$ for $\mathbf{x}[n]$.
These parameters are given by the subsequent equations
\begin{align}
&\frac{A_n}{A^{max}_n}
=\sin\bigg(\frac{\pi \tau}{T_c}\bigg),\\
&
\frac{\pi( 2t_o+\tau )}{T_c}
=
\phi_n + 2k\pi, 0 \leq t_o < T_c,
k\in \mathbf{Z},
\end{align}
where 
$A^{max}_n$ 
denotes the maximum amplitude among all symbols of $\mathbf{x}[n]$
and the matrix $\mathbf{Z}$ is the set of positive integers.
It is worth emphasizing that the primary purpose of these parameters is to 
demonstrate the TMA mechanism used to generate downlink control signals.
The triplet (i.e., $T_c$, $\tau_n$ and $t_{o,n}$) is specified to describe the timing and duration rules 
that underpin control-signal construction,
and they serve as descriptive artifacts for the implementation process. 
Provided these parameters obey the constraints and functional relations presented in this paper, 
varying their specific numerical values will not materially influence or constrain the subsequent beamforming design steps.

The TRIS transceiver design comprises a TRIS formed by reconfigurable elements placed in a uniform planar array (UPA), 
a feed horn antenna and a control unit.
As described in \cite{ref_TRIS_1},
the TRIS transceiver architecture constitutes a power-conscious multi-stream communications platform 
that is particularly advantageous for time-varying propagation environments 
and for deployments where energy resources are constrained.
Based on the novel signal generation mechanism of the TRIS transceiver \cite{ref_TRIS_1},
the beamforming vectors $\mathbf{f}_{I}$ and $\mathbf{f}_{E}$ must satisfy the individual power constraint for each TRIS transceiver antenna,
and the power constraint can be represented as
\begin{align}
\mathbf{f}_{I}^H \bar{\mathbf{A}}_{I,n} \mathbf{f}_{I} 
+ \mathbf{f}_{E}^H \bar{\mathbf{A}}_{E,n}\mathbf{f}_{E} \leq P_t, \forall n \in \mathcal{N},
\end{align}
where
$\mathbf{f}_{I}\triangleq [\mathbf{f}_{I,1}^T,\cdots,\mathbf{f}_{I,K}^T]^T \in \mathbb{C}^{NK \times 1}  $,
$\mathbf{f}_{E}\triangleq [\mathbf{f}_{E,1}^T,\cdots,\mathbf{f}_{E,G}^T]^T \in \mathbb{C}^{NG \times 1}  $,
and
$P_t$ is the maximum power limit for transmission at each TRIS transceiver antenna. 
The selection matrices $\bar{\mathbf{A}}_{I,n}$ and  $\bar{\mathbf{A}}_{E,n}$ are given as
\begin{align}
&\bar{\mathbf{A}}_{I,n} \triangleq \text{blkdiag}(\underbrace{\mathbf{A}_n, \cdots,\mathbf{A}_n}
\limits_{K}
  ) \in \mathbb{R}^{NK\times NK},\\
&\bar{\mathbf{A}}_{E,n} \triangleq \text{blkdiag}(\underbrace{\mathbf{A}_n, \cdots,\mathbf{A}_n}
\limits_{G}
  ) \in \mathbb{R}^{NG\times NG},
\end{align}
respectively,
where
$\mathbf{A}_n \triangleq \text{diag}(\mathbf{a}_{n}) $
and the index vector $\mathbf{a}_{n}$ is denoted as
\begin{align}
\mathbf{a}_{n} \triangleq [0,0,\cdots,\underbrace{1}\limits_{\textrm{n-}th},\cdots,0,0] \in \mathbb{R}^{N\times N},
\end{align}
where the element at position 
$n$ equals $1$ and all other elements are $0$.

Then
the received signal at the $k$-th ID user can be formulated as
\begin{align}
{y}_{I,k}
&=
\underbrace{{\mathbf{h}}_{I,k}^H\mathbf{f}_{I,k}x_{I,k}}
\limits_{\textrm{Desired signal}}
+\underbrace{\sum_{i\neq k}^{K}{\mathbf{h}}_{I,k}^H\mathbf{f}_{I,i}x_{I,i}}
\limits_{\textrm{Other ID users' interference}} \\
&+ \underbrace{\sum_{g=1}^{G} {\mathbf{h}}_{I,k}^H\mathbf{f}_{E,g}x_{E,g}}
\limits_{\textrm{EH users' interference}} 
 + n_{I,k},\nonumber
\end{align}
where 
$n_{I,k} \sim \mathcal{CN}(0, \sigma_{I,k}^2)  $ denotes the complex additive white Gaussian noise (AWGN) of the $k$-th ID user.
For characterizing the performance upper bound of the TRIS transceiver-empowered SWIPT system 
with beamforming design, 
we assume that the TRIS transceiver has the perfect CSI for all relevant channels, 
based on the channel estimation approaches discussed in \cite{ref_channel_estimation_1}$-$\cite{ref_channel_estimation_2}.
The equivalent baseband channels from the TRIS transceiver to the $k$-th ID user and the TRIS transceiver to the $g$-th EH user 
are denoted by
$\mathbf{h}_{I,k} \in \mathbb{C}^{N\times 1} $
and $\mathbf{h}_{E,g}\in \mathbb{C}^{N\times 1}$, 
respectively,
and are assumed to follow the quasi-static flat-fading model.

The total number of TRIS units is expressed as
$N = N_h \times N_v$,
where
$N_h$
and
$N_v$ respectively denotes the number of TRIS units 
in the horizontal and vertical directions.
We assume that the TRIS-ID user $k$ channel is the Rician channel model,
which is formulated as
\begin{align}
\mathbf{h}_{I,k}
=
\sqrt{C_0 \bigg (\frac{d}{d_0} \bigg)^{-\alpha}}
\bigg(
\sqrt{\frac{\kappa}{\kappa+1}}\mathbf{h}_{I,k}^{LoS}
+
\sqrt{\frac{1}{\kappa+1}}\mathbf{h}_{I,k}^{NLoS}
\bigg),
\end{align}
where
$C_0$ denotes as the large-scale fading coefficient with reference distance $d_0 = 1$m.  
$d$ and $\alpha$ represent the distance and the large-scale fading factor of the corresponding channel, 
respectively.
$\kappa$ is the Rician factor.
Additionally, 
the term $\mathbf{h}_{I,k}^{LoS}$ 
is the line-of-sight (LoS) component of 
the channel bwtween the TRIS transceiver and the $k$-th ID user,
which is given in (\ref{h_LoS}),
\begin{figure*}
\begin{align}
\mathbf{h}_{I,k}^{LoS}
&=
\bigg[
1,
e^{-j \frac{2\pi}{\lambda} d \sin \theta_{I,k}^{AoD} \cos\psi_{I,k}^{AoD}  },
\cdots,
e^{-j \frac{2\pi}{\lambda} (N_h-1) d \sin \theta_{I,k}^{AoD} \cos\psi_{I,k}^{AoD}  }
\bigg]^T \label{h_LoS} \\
& \otimes\bigg[
1,
e^{-j \frac{2\pi}{\lambda} d \sin \theta_{I,k}^{AoD} \cos\psi_{I,k}^{AoD}  },
\cdots,
e^{-j \frac{2\pi}{\lambda} (N_z-1) d \sin \theta_{I,k}^{AoD} \cos\psi_{I,k}^{AoD}  }
\bigg]^T,\nonumber
\end{align}
\boldsymbol{\hrule}
\end{figure*}
where
$\theta_{I,k}^{AoD}$
and
$\psi_{I,k}^{AoD}$
respectively represent the vertical and horizontal angle-of-departure (AoD) at the TRIS transceiver,
$d$ is the spacing between the transmissive units of the TRIS transceiver,
and $\lambda$ denotes the wavelength of the carrier.
The nonline-of-sight (NLoS) component of the channel between the RIS transceiver and the $k$-th ID user 
is represented as
$\mathbf{h}_{I,k}^{NLoS}$,
where
$\mathbf{h}_{I,k}^{NLoS}[(n_h-1)N_v + n_v] \sim\mathcal{CN}(0,1)$ 
is the $(n_h-1)N_v + n_v$-the unit of $\mathbf{h}_{I,k}^{NLoS}$.
Furthermore,
the TRIS-EH user $g$ channel is also assumed as the Rician channel model.
The channel model for the TRIS-EH user $g$ link is omitted here to avoid repetition.

Let 
$ \bar{\mathbf{h}}_{I,k} \triangleq [ \mathbf{h}_{I,k}^T,\cdots, \mathbf{h}_{I,k}^T ]^T \in \mathbb{C}^{NK \times 1}  $
and
$ \bar{\mathbf{h}}_{I2,k} \triangleq [ \mathbf{h}_{I,k}^T,\cdots, \mathbf{h}_{I,k}^T ]^T \in \mathbb{C}^{NG \times 1}  $.
The received signal ${y}_{I,k}$ can be rewritten as
\begin{align}
{y}_{I,k}
&=
\underbrace{\bar{\mathbf{h}}_{I,k}^H\mathbf{B}_{I,k}\mathbf{f}_{I}x_{I,k}}
\limits_{\textrm{Desired signal}}
+\underbrace{\sum_{i\neq k}^{K}\bar{\mathbf{h}}_{I,k}^H\mathbf{B}_{I,i}\mathbf{f}_{I}x_{I,i}}
\limits_{\textrm{Other ID users' interference}} \\
&+ \underbrace{\sum_{g=1}^{G} \bar{\mathbf{h}}_{I2,k}^H\mathbf{B}_{E,g}\mathbf{f}_{E}x_{E,g}}
\limits_{\textrm{EH users' interference}} 
 + n_{I,k},\nonumber
\end{align}
where 
$\mathbf{B}_{I,k} \triangleq \text{diag}(\mathbf{b}_{I,k}) \in \mathbb{R}^{NK\times NK}$
and
$\mathbf{B}_{E,g} \triangleq \text{diag}(\mathbf{b}_{E,g}) \in \mathbb{R}^{NG\times NG}$
are the selection matrices,
and 
the vector
$\mathbf{b}_{I,k}$ 
is  defined as
\begin{align}
&\mathbf{b}_{I,k} \triangleq [0,\cdots,0,\underbrace{1,\cdots,1}
\limits_{N}
,0,\cdots,0]^T \in \mathbb{R}^{NK\times 1},
\end{align}
with the entries in the index range
$((k-1)\times N +1)$ to $ (k\times N )$
set to 1, 
and all other entries equal to zero. 
Similarly, 
the vector 
$\mathbf{b}_{E,g}$ 
is defined in the same way as follows
\begin{align}
&\mathbf{b}_{E,g} \triangleq [0,\cdots,0,\underbrace{1,\cdots,1}
\limits_{N}
,0,\cdots,0]^T \in \mathbb{R}^{NG\times 1},
\end{align}
where the entries in the index range
$((g-1)\times N +1)$ to $ (g\times N )$
are
set to 1, 
and all other entries equal to zero.

Next,
the received signal at the $g$-th EH user can be expressed as
\begin{align}
{y}_{E,g}
&=
\sum_{i=1}^{G} {\mathbf{h}}_{E,g}^H\mathbf{f}_{E,i}x_{E,i}
+
\sum_{k=1}^{K}{\mathbf{h}}_{E,g}^H\mathbf{f}_{I,k}x_{I,k}
+ n_{E,g},\\
&=
\sum_{i=1}^{G}\bar{\mathbf{h}}_{E,g}^H \mathbf{B}_{E,i}  \mathbf{f}_{E}x_{E,i}
+\sum_{k=1}^{K}\bar{\mathbf{h}}_{E2,g}^H\mathbf{B}_{I,k} \mathbf{f}_{I}x_{I,k}
+ n_{E,g},\nonumber
\end{align}
where 
$n_{E,g} \sim \mathcal{CN}(0, \sigma_{E,g}^2)  $ is the complex AWGN of the $g$-th EH user,
$ \bar{\mathbf{h}}_{E,g} \triangleq [ \mathbf{h}_{E,g}^T,\cdots, \mathbf{h}_{E,g}^T ]^T \in \mathbb{C}^{NG \times 1}  $,
and
$ \bar{\mathbf{h}}_{E2,g} \triangleq [ \mathbf{h}_{E,g}^T,\cdots, \mathbf{h}_{E,g}^T ]^T \in \mathbb{C}^{NK \times 1}  $.

\subsection{Performance Metrics}

The received SINR at the $k$-th ID user is given by
\begin{align}
&\text{SINR}_{k}( \mathbf{f}_{I}, \mathbf{f}_{E}  )\\
&=
\frac{\vert\bar{\mathbf{h}}_{I,k}^H\mathbf{B}_{I,k}\mathbf{f}_{I}  \vert^2}
{
\sum_{i\neq k}^{K}\vert \bar{\mathbf{h}}_{I,k}^H\mathbf{B}_{I,i}\mathbf{f}_{I}\vert^2
+
\sum_{g=1}^{G} \vert \bar{\mathbf{h}}_{I2,k}^H\mathbf{B}_{E,g}\mathbf{f}_{E}\vert^2
+
\sigma_{I,k}^2
}.\nonumber
\end{align}
Therefore,
the corresponding achievable rate of the $k$-th ID user is formulated as
\begin{align}
\mathrm{R}_{k}( \mathbf{f}_{I}, \mathbf{f}_{E}  ) =\text{log}\bigg( 1 + \text{SINR}_{k}( \mathbf{f}_{I}, \mathbf{f}_{E}  )\bigg).
\end{align}

On the other hand,
since EH users can harvest wireless energy form both the information and energy signals,
we assume that the EH users adopt a linear EH model \cite{ref_RIS_SWIPT_5}$-$\cite{ref_RIS_SWIPT_6}.
Since the noise power is much smaller than the power received
from the TRIS transceiver,
the energy harvested from the noise is assumed to be negligible.
Thus,
the harvested energy at the $g$-th EH user 
is represented as
\begin{align}
\mathrm{Q}_{g}( \mathbf{f}_{I}, \mathbf{f}_{E}  )
&=\zeta
\bigg(
\vert\bar{\mathbf{h}}_{E,g}^H\mathbf{B}_{E,g}\mathbf{f}_{E} \vert^2\\
&+
\underbrace{\sum_{i\neq g}^{G}
\vert\bar{\mathbf{h}}_{E,g}^H\mathbf{B}_{E,i}\mathbf{f}_{E} \vert^2}
\limits_{\textrm{Harvested energy from other EH signals}}
\nonumber\\
&+
\underbrace{
\sum_{k=1}^{K}\vert\bar{\mathbf{h}}_{E2,g}^H\mathbf{B}_{I,k}\mathbf{f}_{I} \vert^2}
\limits_{\textrm{Harvested energy from ID signals}}
\bigg),\nonumber
\end{align}
where
$\bar{\mathbf{h}}_{E,g} \triangleq [\mathbf{h}_{E,g}^T, \cdots ,\mathbf{h}_{E,g}^T  ]^T \in \mathbb{C}^{NG \times 1} $
and
$\bar{\mathbf{h}}_{E2,g} \triangleq [\mathbf{h}_{E,g}^T, \cdots ,\mathbf{h}_{E,g}^T  ]^T \in \mathbb{C}^{NK \times 1} $,
$0< \zeta \leq 1$ is the energy harvesting efficiency.

\subsection{Problem Formulation}

In this paper,
our objective is to maximize the weighted sum-rate of all ID users by optimizing the active beamforming of the TRIS transceiver,
subject to a total harvested energy constraint of all EH users and 
the per-element transmit power constraints at the TRIS transceiver units.
Thus,
the corresponding optimization problem is mathematically formulated as
\begin{subequations}
\begin{align}
\textrm{(P0)}:&\mathop{\textrm{max}}
\limits_{\mathbf{f}_I, \mathbf{f}_E
}\
\sum_{k=1}^{K}
\omega_{ID,k}
\mathrm{R}_{k}(\mathbf{f}_I, \mathbf{f}_E) 
\label{P0_obj}\\
\textrm{s.t.}\ 
& \sum_{g=1}^{G}  \mathrm{Q}_g(\mathbf{f}_I, \mathbf{f}_E)  \geq Q_t,\label{P0_c_0}\\
& \mathbf{f}_{I}^H\mathbf{\bar{A}}_{I,n}\mathbf{f}_{I} + \mathbf{f}_{E}^H\mathbf{\bar{A}}_{E,n}\mathbf{f}_{E} \leq P_t, \forall n \in \mathcal{N},\label{P0_c_1}
\end{align}
\end{subequations}
where
$\omega_{ID,k}$ denotes the weight of the $k$-th ID user.
To ensure the fairness among all ID users,
the SWIPT system designer can able to impose desired priority levels by adjust the values of the weights.
Given that the weights do not affect the algorithm effect,
we assume that all ID users share the same weight for simplicity and without loss of generality,
i.e., $\omega_{ID,k} = 1, \forall k \in \mathcal{K}$.
(\ref{P0_c_0}) indicates the harvested energy constraint for all
EH users with $Q_t$ denoting the minimum total harvested
energy; (\ref{P0_c_1}) is the transmit power constraint for the TRIS transceiver unit.

The problem (P0) is highly challenging to tackle 
due to its non-convex objective function as well as non-convex harvested energy constraint.
Besides, accurately representing the limited amplification potential of individual TRIS units 
requires imposing power restrictions for each element separately.
Adopting this element-wise constraint approach injects a significant number of new inequalities into the optimization formulation, 
since every antenna unit is associated with its own allowable power range.
As a result, 
the problem's constraint count grows considerably and the optimization landscape becomes markedly more difficult, 
both analytically and numerically, 
to handle.
In general,
there are no standard algorithms for solving the above non-convex optimization problem optimally.
Therefore,
we will firstly propose an SOCP-based algorithm to solve (P0),
and then consider a low-complexity solution to solve (P0) for reducing the computation complexity.

\section{SOCP-based Algorithm}

In this section,
we propose an SOCP-based algorithm  to solve the original optimization problem (P0).
Specifically,
the original optimization problem is firstly transformed into an equivalent optimization problem.
And then, 
the block coordinate ascent (BCA) method \cite{ref_BCA} is employed to alternately update the 
auxiliary and primal variables.

\subsection{Problem Reformulation}
First, 
the one difficulty for solving the problem (P0) lies in the sum-rate objective function.
To enhance the tractability of problem (P0), 
we employ the WMMSE framework \cite{ref_WMMSE} to reformulate the original objective function
$\mathrm{R}_{k}(\mathbf{f}_I, \mathbf{f}_E) $
 into a more tractable formulation.
In particular, 
with the introduction of auxiliary variables $\{\beta_k \}$ and $\{\omega_k \}$,
the function 
$\mathrm{R}_{k}(\mathbf{f}_I, \mathbf{f}_E) $ 
can be transformed into an equivalent variational form presented in (\ref{WMMSE_transformation}).

\begin{figure*}
\begin{align}
&\mathrm{R}_{k}(\mathbf{f}_I, \mathbf{f}_E)
=\text{log}
\bigg(
1
+
\text{SINR}_{k}( \mathbf{f}_{I}, \mathbf{f}_{E}  )
\bigg)\label{WMMSE_transformation}
\\
&=
\text{log}
\bigg(
1
+
\frac{\vert\bar{\mathbf{h}}_{I,k}^H\mathbf{B}_{I,k}\mathbf{f}_{I}  \vert^2}
{
{\sum}_{i\neq k}^{K}\vert \bar{\mathbf{h}}_{I,k}^H\mathbf{B}_{I,i}\mathbf{f}_{I}\vert^2
+
{\sum}_{g=1}^{G} \vert \bar{\mathbf{h}}_{I2,k}^H\mathbf{B}_{E,g}\mathbf{f}_{E}\vert^2
+
\sigma_{I,k}^2
}
\bigg)\nonumber\\
&=\mathop{\textrm{max}}
\limits_{
\omega_k\geq0
}
\bigg\{
\textrm{log}(\omega_k)-\omega_k\big( 
{\sum}_{i=1}^{K}\vert \bar{\mathbf{h}}_{I,k}^H\mathbf{B}_{I,i}\mathbf{f}_{I}\vert^2
+
{\sum}_{g=1}^{G} \vert \bar{\mathbf{h}}_{I2,k}^H\mathbf{B}_{E,g}\mathbf{f}_{E}\vert^2
+
\sigma_{I,k}^2
\big)^{-1}\bar{\mathbf{h}}_{I,k}^H\mathbf{B}_{I,k}\mathbf{f}_{I} + 1
\bigg\}
\nonumber\\
&= 
\mathop{\textrm{max}}
\limits_{
\omega_k\geq0,
\beta_k
}
\bigg\{
\underbrace{
\textrm{log}(\omega_k)-\omega_k\big( 
1-2\text{Re}\{ \beta_k^{\ast} \bar{\mathbf{h}}_{I,k}^H\mathbf{B}_{I,k}\mathbf{f}_{I}\}
+ \vert\beta_k\vert^2(
{\sum}_{i=1}^{K}\vert \bar{\mathbf{h}}_{I,k}^H\mathbf{B}_{I,i}\mathbf{f}_{I}\vert^2
+
{\sum}_{g=1}^{G} \vert \bar{\mathbf{h}}_{I2,k}^H\mathbf{B}_{E,g}\mathbf{f}_{E}\vert^2
+
\sigma_{I,k}^2)
\big) + 1}
\limits_{\mathrm{\tilde{R}}_k(\mathbf{f}_I, \mathbf{f}_E,\omega_k,\beta_k)} \bigg\}.
 \nonumber
\end{align}
\boldsymbol{\hrule}
\end{figure*}

Consequently, 
the original problem (P0) can be equivalently represented by
\begin{subequations}
\begin{align}
\textrm{(P1)}:&\mathop{\textrm{max}}
\limits_{\mathbf{f}_I, \mathbf{f}_E, \{\omega_k\}, \{\beta_k\}
}\
\sum_{k=1}^{K}
\mathrm{\tilde{R}}_k(\mathbf{f}_I, \mathbf{f}_E,\omega_k,\beta_k)
\label{P1_obj}\\
\textrm{s.t.}\ 
& \sum_{g=1}^{G}  \mathrm{Q}_g(\mathbf{f}_I, \mathbf{f}_E)  \geq Q_t,\label{P1_c_0}\\
& \mathbf{f}_{I}^H\mathbf{\bar{A}}_{I,n}\mathbf{f}_{I} + \mathbf{f}_{E}^H\mathbf{\bar{A}}_{E,n}\mathbf{f}_{E} 
\leq P_t, \forall n \in \mathcal{N}.\label{P1_c_1}
\end{align}
\end{subequations}

Note that the optimization variables 
(i.e.,
$\mathbf{f}_I$
and 
$\mathbf{f}_E$)
and 
the auxiliary variables
(i.e.,
$\{\omega_k\}$
and 
$\{\beta_k\}$)
are coupled in the objective function (\ref{P6_obj}) of problem (P1),
which makes the problem challenging to solve directly.
Therefore, 
the BCA method \cite{ref_BCA} is adopted to efficiently solve problem (P1).
Specifically, 
we partition all the optimization variables in problem (P1) into three blocks, 
i.e., $\{\omega_k\}$, 
$\{ \beta_k\}$ 
and 
$\{ \mathbf{f}_I,
\mathbf{f}_E \}$.
The objective in (P1) is maximized by alternatingly optimizing one block of variables 
while keeping the other blocks fixed, 
and repeating this procedure until convergence.
The updated details are presented in following 
and the convergence is achieved when the fractional decrease of the objective
function is less than a sufficiently small threshold or when a maximum number of iterations is reached.

\subsection{Optimizing auxiliary variables}

Based on the WMMSE transformation, 
with all other variables fixed, 
the auxiliary variables 
$\{\beta_k\}$
and
$\{\omega_k\}$
can be straightforwardly updated in closed form, which are respectively given as follows 
\begin{align}
&  \beta_k^{\star}
=
\frac{\bar{\mathbf{h}}_{I,k}^H\mathbf{B}_{I,k}\mathbf{f}_{I}}
{\sum_{i=1}^{K}\vert \bar{\mathbf{h}}_{I,k}^H\mathbf{B}_{I,i}\mathbf{f}_{I}\vert^2
+
\sum_{g=1}^{G} \vert \bar{\mathbf{h}}_{I2,k}^H\mathbf{B}_{E,g}\mathbf{f}_{E}\vert^2
+
\sigma_{I,k}^2}, \label{beta_opt}\\
&\omega_k^{\star}
=
1\!+\!
\frac{\bar{\mathbf{h}}_{I,k}^H\mathbf{B}_{I,k}\mathbf{f}_{I}}
{\sum_{i \neq k}^{K}\vert \bar{\mathbf{h}}_{I,k}^H\mathbf{B}_{I,i}\mathbf{f}_{I}\vert^2
\!+\!
\sum_{g=1}^{G} \vert \bar{\mathbf{h}}_{I2,k}^H\mathbf{B}_{E,g}\mathbf{f}_{E}\vert^2
\!+\!
\sigma_{I,k}^2}.\label{omega_opt}
\end{align}

\subsection{Updating The Beamformer}
In the following subsection,
we will present the procedure for updating the transmit beamforming when other variables are fixed.
First,
the objective function
$\mathrm{\tilde{R}}_k(\mathbf{f}_I, \mathbf{f}_E,\omega_k,\beta_k)$
of problem (P1) can be equivalently rewritten as
\begin{align}
&\mathrm{\tilde{R}}_k(\mathbf{f}_I, \mathbf{f}_E,\omega_k,\beta_k)\\
&=
-\mathbf{f}_I^H \mathbf{B}_{1,k} \mathbf{f}_I
-
\mathbf{f}_E^H\mathbf{B}_{2,k}\mathbf{f}_E 
+
 2\text{Re}\{ \mathbf{b}_{1,k}^H \mathbf{f}_I \} + c_{1,k},\nonumber
\end{align}
where the newly introduced coefficients are given as
\begin{align}
&\mathbf{B}_{1,k}  \triangleq   
\sum_{i=1}^{K} 
\omega_k\vert \beta_k \vert^2
(
\mathbf{B}_{I,i} \mathbf{\bar{h}}_{I,k}\mathbf{\bar{h}}_{I,k}^H \mathbf{B}_{I,i}
)
,\\
& \mathbf{B}_{2,k} \triangleq   
\sum_{g=1}^{G} 
\omega_k\vert \beta_k \vert^2
(
\mathbf{B}_{E,g} \mathbf{\bar{h}}_{I2,k}\mathbf{\bar{h}}_{I2,k}^H \mathbf{B}_{E,g}
), \nonumber\\
&c_{1,k}
\triangleq
\text{log}(\omega_k) - \omega_k
- \omega_k\vert \beta_k \vert^2  \sigma_{I,k}^2  + 1,\nonumber\\
& \mathbf{b}_{1,k} \triangleq   
\omega_k\beta_k \mathbf{B}_{I,k}\mathbf{\bar{h}}_{I,k}.\nonumber
\end{align}

Therefore,
the sum-rate function
${\sum}_{k=1}^{K}
\mathrm{\tilde{R}}_k(\mathbf{f}_I, \mathbf{f}_E,\omega_k,\beta_k)$
in the problem (P1)
 can be rewritten 
as follows
\begin{align}
&\sum_{k=1}^{K}\mathrm{\tilde{R}}_k(\mathbf{f}_I, \mathbf{f}_E,\omega_k,\beta_k)\\
&=
-\mathbf{f}_I^H \mathbf{B}_{3} \mathbf{f}_I
-
\mathbf{f}_E^H\mathbf{B}_{4}\mathbf{f}_E 
+
 2\text{Re}\{ \mathbf{b}_{2}^H \mathbf{f}_I \} + c_{2},\nonumber
\end{align}
where
\begin{align}
&\mathbf{B}_{3}  \triangleq   
\sum_{k=1}^{K} 
\mathbf{B}_{1,k},
 \mathbf{B}_{4} \triangleq   
\sum_{k=1}^{K} 
\mathbf{B}_{2,k}, \\
&c_{2}
\triangleq
\sum_{k=1}^{K} 
c_{1,k},
 \mathbf{b}_{2} \triangleq   
\sum_{k=1}^{K} \mathbf{b}_{1,k}.\nonumber
\end{align}

Next,
the harvested energy function 
$\mathrm{Q}_g(\mathbf{f}_I, \mathbf{f}_E)$ in problem (P1),
corresponding to the $g$-th EH user,
can be reformulated in the following form
\begin{align}
\mathrm{Q}_g(\mathbf{f}_I, \mathbf{f}_E)
=
\mathbf{f}_I^H \mathbf{B}_{5,g} \mathbf{f}_I
+
\mathbf{f}_E^H\mathbf{B}_{6,g}\mathbf{f}_E,
\end{align}
where
\begin{align}
&\mathbf{B}_{5,g}  \triangleq   
\sum_{k=1}^{K} 
\mathbf{B}_{I,k}\bar{\mathbf{h}}_{E2,g}\bar{\mathbf{h}}_{E2,g}^H\mathbf{B}_{I,k}
,\\
& \mathbf{B}_{6,g} \triangleq   
\sum_{i=1}^{G} 
\mathbf{B}_{E,i}\bar{\mathbf{h}}_{E,g}\bar{\mathbf{h}}_{E,g}^H\mathbf{B}_{E,i}.
\nonumber
\end{align}

And then,
the total harvested energy constraint (\ref{P1_c_1})
can be rewritten as follows
\begin{align}
&\sum_{g=1}^{G}  \mathrm{Q}_g(\mathbf{f}_I, \mathbf{f}_E)  \geq Q_t
\\
&
\Longleftrightarrow
\mathbf{f}_I^H \mathbf{B}_{7} \mathbf{f}_I
+
\mathbf{f}_E^H\mathbf{B}_{8}\mathbf{f}_E
\geq Q_t
,\nonumber
\end{align}
where
\begin{align}
&\mathbf{B}_{7}  \triangleq   
\sum_{g=1}^{K} 
\mathbf{B}_{5,g}
, \mathbf{B}_{8} \triangleq   
\sum_{g=1}^{G} 
\mathbf{B}_{6,g}.
\end{align}

Based the above transformations,
the problem (P1) can be recasted as
\begin{subequations}
\begin{align}
\textrm{(P2)}:\mathop{\textrm{max}}
\limits_{\mathbf{f}_I, \mathbf{f}_E
}\
&-\mathbf{f}_I^H \mathbf{B}_{3} \mathbf{f}_I
-
\mathbf{f}_E^H\mathbf{B}_{4}\mathbf{f}_E 
+
 2\text{Re}\{ \mathbf{b}_{2}^H \mathbf{f}_I \} + c_{2}\label{P2_obj} 
\\
\textrm{s.t.}\ 
& \mathbf{f}_I^H \mathbf{B}_{7} \mathbf{f}_I
+
\mathbf{f}_E^H\mathbf{B}_{8}\mathbf{f}_E  \geq Q_t,\label{P2_c_0}\\
& \mathbf{f}_{I}^H\mathbf{\bar{A}}_{I,n}\mathbf{f}_{I} + \mathbf{f}_{E}^H\mathbf{\bar{A}}_{E,n}\mathbf{f}_{E} 
\leq P_t, \forall n \in \mathcal{N}.\label{P2_c_1}
\end{align}
\end{subequations}

Obviously, 
the problem (P2) is non-convex since the constraint (\ref{P2_c_0}) is non-convex.
Therefore, 
following the MM framework \cite{ref_MM}, 
we linearize
the quadratic terms
$\mathbf{f}_I^H \mathbf{B}_{7} \mathbf{f}_I$
and
$\mathbf{f}_E^H\mathbf{B}_{8}\mathbf{f}_E$, respectively, 
to obtain the following tight lower bounds
\begin{align}
&\mathbf{f}_I^H \mathbf{B}_{7} \mathbf{f}_I
\geq
\mathbf{f}_{I,0}^H \mathbf{B}_{7} \mathbf{f}_{I,0}
+
2\text{Re}\{\mathbf{f}_{I,0}^H \mathbf{B}_{7}( \mathbf{f}_{I}-\mathbf{f}_{I,0} )  \}
, \label{P2_MM_1}\\
&\mathbf{f}_E^H\mathbf{B}_{8}\mathbf{f}_E
\geq
\mathbf{f}_{E,0}^H \mathbf{B}_{8} \mathbf{f}_{E,0}
+
2\text{Re}\{\mathbf{f}_{E,0}^H \mathbf{B}_{8}( \mathbf{f}_{E}-\mathbf{f}_{E,0} )  \},\label{P2_MM_2}
\end{align}
where 
$\mathbf{f}_{I,0}$
and
$\mathbf{f}_{E,0}$
 are the values obtained in the last iteration.

Therefore, 
by replace the quadratic terms
$\mathbf{f}_I^H \mathbf{B}_{7} \mathbf{f}_I$
and
$\mathbf{f}_E^H\mathbf{B}_{8}\mathbf{f}_E$ by 
(\ref{P2_MM_1})
and
(\ref{P2_MM_2}), respectively,
the problem (P2) 
can be represented as
\begin{subequations}
\begin{align}
\textrm{(P3)}:&\mathop{\textrm{max}}
\limits_{\mathbf{f}_I, \mathbf{f}_E
}\
-\mathbf{f}_I^H \mathbf{B}_{3} \mathbf{f}_I
-
\mathbf{f}_E^H\mathbf{B}_{4}\mathbf{f}_E +
 2\text{Re}\{ \mathbf{b}_{2}^H \mathbf{f}_I \} + c_{2}
\label{P3_obj}\\
\textrm{s.t.}\ 
& \mathbf{f}_{I,0}^H \mathbf{B}_{7} \mathbf{f}_{I,0}
+
2\text{Re}\{\mathbf{f}_{I,0}^H \mathbf{B}_{7}( \mathbf{f}_{I}-\mathbf{f}_{I,0} )  \}\label{P3_c_0}\\
&+
\mathbf{f}_{E,0}^H \mathbf{B}_{8} \mathbf{f}_{E,0}
+
2\text{Re}\{\mathbf{f}_{E,0}^H \mathbf{B}_{8}( \mathbf{f}_{E}-\mathbf{f}_{E,0} )  \}  \geq Q_t, \nonumber\\
& \mathbf{f}_{I}^H\mathbf{\bar{A}}_{I,n}\mathbf{f}_{I} + \mathbf{f}_{E}^H\mathbf{\bar{A}}_{E,n}\mathbf{f}_{E} 
\leq P_t, \forall n \in \mathcal{N}.\label{P3_c_1}
\end{align}
\end{subequations}

The problem (P3) is a typical SOCP
and can be solved by CVX \cite{ref_CVX}.
The SOCP-based method can be summarized in Algorithm \ref{alg:2}.

\begin{algorithm}[t]
\caption{The SOCP-based Method}
\label{alg:2}
\begin{algorithmic}[1]
\STATE {initialize}
$\mathbf{f}_I^{(0)}$,
$\mathbf{f}_E^{(0)}$
and
$t=0$
;
\REPEAT
\STATE update $\{\beta_k^{(t+1)}\}$ and $\{\omega_k^{(t+1)}\}$ by (\ref{beta_opt}) and (\ref{omega_opt}), respectively;
\STATE update $\mathbf{f}_I^{(t+1)}$ and $\mathbf{f}_E^{(t+1)}$ by solving  (P3);
\STATE $t++$;
\UNTIL{$convergence$;}
\end{algorithmic}
\end{algorithm}

\section{Low-complexity Algorithm}

It is worth mentioning that
 our previously proposed SOCP-based algorithm (i.e.,  Alg. \ref{alg:2})  depends on numerical solvers, e.g., CVX, 
to update the beamforming
(i.e., $\mathbf{f}_I$
and
$\mathbf{f}_E$).
However,
since general convex optimization solvers such as CVX employs interior-point (IP) method \cite{ref_Convex Optimization} 
to solve SOCP problem,
the complexity of solving SOCP problem increases dramatically as the dimensionality of the variables grows.
Therefore, 
we turn to the design of a low-complexity method that avoids reliance on any numerical optimization solvers.

\subsection{Efficient Update of The Beamformer}

First,
we rewrite the objective function (\ref{P3_obj}) as follows
\begin{align}
&-\mathbf{f}_I^H \mathbf{B}_{3} \mathbf{f}_I
-
\mathbf{f}_E^H\mathbf{B}_{4}\mathbf{f}_E 
+
 2\text{Re}\{ \mathbf{b}_{2}^H \mathbf{f}_I \} + c_{2}\\
& \Leftrightarrow
-\mathbf{f}_{IE}^H \mathbf{B}_{9} \mathbf{f}_{IE}
+
 2\text{Re}\{ \mathbf{b}_{3}^H \mathbf{f}_{IE} \} + c_{2},
 \nonumber
\end{align}
where
\begin{align}
&\mathbf{f}_{IE} \triangleq [\mathbf{f}_{I}^T, \mathbf{f}_{E}^T]^T,
\mathbf{b}_3 \triangleq [ \mathbf{b}_2^T , \mathbf{0}^T]^T,\\
&\mathbf{B}_9 \triangleq \text{blkdiag}( \mathbf{B}_3, \mathbf{B}_4).\nonumber
\end{align}

And then,
the constraints (\ref{P3_c_0}) and (\ref{P3_c_1}) can be respectively rewritten as
\begin{align}
&\mathbf{f}_{I,0}^H \mathbf{B}_{7} \mathbf{f}_{I,0}
+
2\text{Re}\{\mathbf{f}_{I,0}^H \mathbf{B}_{7}( \mathbf{f}_{I}-\mathbf{f}_{I,0} )  \}\\
&+
\mathbf{f}_{E,0}^H \mathbf{B}_{8} \mathbf{f}_{E,0}
+
2\text{Re}\{\mathbf{f}_{E,0}^H \mathbf{B}_{8}( \mathbf{f}_{E}-\mathbf{f}_{E,0} )  \}  \geq Q_t\nonumber\\
&
\Leftrightarrow
- 2\text{Re}\{ \mathbf{b}_{4}^H \mathbf{f}_{IE} \} - c_{3} \leq 0, \nonumber \\
& \mathbf{f}_{I}^H\mathbf{\bar{A}}_{I,n}\mathbf{f}_{I} + \mathbf{f}_{E}^H\mathbf{\bar{A}}_{E,n}\mathbf{f}_{E} \leq P_t\\
&\Leftrightarrow
\mathbf{f}_{IE}^H\mathbf{\bar{A}}_{IE,n}\mathbf{f}_{IE} \leq P_t,\nonumber
\end{align}
where
\begin{align}
&\mathbf{B}_{10} \triangleq \text{blkdiag}( \mathbf{B}_7, \mathbf{B}_8), \\ 
&\mathbf{\bar{A}}_{IE,n} \triangleq \text{blkdiag}( \mathbf{\bar{A}}_{I,n}, \mathbf{\bar{A}}_{E,n}),\nonumber\\
&\mathbf{f}_{IE,0} \triangleq [\mathbf{f}_{I,0}^T, \mathbf{f}_{E,0}^T]^T,
 \mathbf{b}_4 \triangleq \mathbf{B}_{10}^H \mathbf{f}_{IE,0},\nonumber\\
&c_{3} \triangleq - ( \mathbf{f}_{IE,0}^H\mathbf{\bar{A}}_{IE,n}\mathbf{f}_{IE,0} )^{\ast} - Q_t. \nonumber
\end{align}

Based on the above transformation, 
the optimization problem (P3) reduces to solving the following problem
\begin{subequations}
\begin{align}
\textrm{(P4)}:&\mathop{\textrm{min}}
\limits_{\mathbf{f}_{IE}
}\
\mathbf{f}_{IE}^H \mathbf{B}_{9} \mathbf{f}_{IE}
-
 2\text{Re}\{ \mathbf{b}_{3}^H \mathbf{f}_{IE} \} - c_{2}
\label{P4_obj}\\
\textrm{s.t.}\ 
& - 2\text{Re}\{ \mathbf{b}_{4}^H \mathbf{f}_{IE} \} - c_{3} \leq 0, \\
& \mathbf{f}_{IE}^H\mathbf{\bar{A}}_{IE,n}\mathbf{f}_{IE} \leq P_t, \forall n \in \mathcal{N}.\label{P4_c_1}
\end{align}
\end{subequations}

Next,
by introduce an auxiliary variable
$\mathbf{w}$,
we can transform (P4) into an equivalent form as follows

\begin{subequations}
\begin{align}
\textrm{(P5)}:&\mathop{\textrm{min}}
\limits_{\mathbf{f}_{IE}, \mathbf{w}
}\
\mathbf{f}_{IE}^H \mathbf{B}_{9} \mathbf{f}_{IE}
-
 2\text{Re}\{ \mathbf{b}_{3}^H \mathbf{f}_{IE} \} - c_{2}
\label{P5_obj}\\
\textrm{s.t.}\ 
& - 2\text{Re}\{ \mathbf{b}_{4}^H \mathbf{f}_{IE} \} - c_{3} \leq 0, \\
& \mathbf{w}^H\mathbf{\bar{A}}_{IE,n}\mathbf{w} \leq P_t, \forall n \in \mathcal{N},\label{P5_c_1}\\
& \mathbf{f}_{IE}=\mathbf{w}.\label{P5_c_2}
\end{align}
\end{subequations}

Subsequently,
we apply the ADMM framework  \cite{ref_RIS_app_5}, \cite{ref_ADMM} to solve the above problem (P5).
To this end, 
by penalizing the equality constraint (\ref{P5_c_2}) in the objective function (\ref{P5_obj}),
the augmented Lagrangian (AL) problem of problem (P5) is formulated as

\begin{subequations}
\begin{align}
\textrm{(P6)}:&\mathop{\textrm{min}}
\limits_{\mathbf{f}_{IE}, \mathbf{w}, \boldsymbol{\tau}
}\
\mathbf{f}_{IE}^H \mathbf{B}_{9} \mathbf{f}_{IE}
-
 2\text{Re}\{ \mathbf{b}_{3}^H \mathbf{f}_{IE} \} - c_{2}\label{P6_obj}\\
 & + \text{Re}\{ \boldsymbol{\tau}^H(\mathbf{f}_{IE} -  \mathbf{w} ) \}
 + \frac{\rho}{2} \Vert\mathbf{f}_{IE} -  \mathbf{w} \Vert_2^2 \nonumber
\\
\textrm{s.t.}\ 
& - 2\text{Re}\{ \mathbf{b}_{4}^H \mathbf{f}_{IE} \} - c_{3} \leq 0, \\
& \mathbf{w}^H\mathbf{\bar{A}}_{IE,n}\mathbf{w} \leq P_t, \forall n \in \mathcal{N},\label{P6_c_1}
\end{align}
\end{subequations}
where $\rho$ is a positive constant
and 
$\boldsymbol{\tau} \in \mathbb{C}^{N(K+G) \times 1}$ is the Lagrangian multiplier.
Under the ADMM methodology,
the problem (P6) can be solved by sequentially updating 
$\mathbf{f}_{IE}$, 
$\mathbf{w}$
and
$\boldsymbol{\tau} $
in an alternative way.
Specifically, 
the objective in AL problem (P6) is minimize by alternatingly optimizing each variable 
while fixing the rest, repeating these updates until convergence. 
The updated details are shown in following.

When $\mathbf{w}$
and
$\boldsymbol{\tau} $
are fixed,
the optimization problem with respective to (w.r.t.) $\mathbf{f}_{IE}$ is given as
\begin{subequations}
\begin{align}
\textrm{(P7)}:&\mathop{\textrm{min}}
\limits_{ \mathbf{f}_{IE}
}\
\mathbf{f}_{IE}^H \bar{\mathbf{B}_{9}} \mathbf{f}_{IE}
-
 2\text{Re}\{ \bar{\mathbf{b}}_{3}^H \mathbf{f}_{IE} \} - \bar{c}_{2}\label{P7_obj}
\\
\textrm{s.t.}\ 
& - 2\text{Re}\{ \mathbf{b}_{4}^H \mathbf{f}_{IE} \} - c_{3} \leq 0,
\end{align}
\end{subequations}
where
\begin{align}
&\bar{\mathbf{B}}_{9} 
\triangleq 
{\mathbf{B}}_{9} + \frac{\rho}{2}\mathbf{I},
\bar{\mathbf{b}}_{3}
\triangleq 
\mathbf{b}_{3} - \frac{1}{2}\boldsymbol{\tau} + \frac{\rho}{2}\mathbf{w},\\
&\bar{c}_{2}
\triangleq 
{c}_{2} + \text{Re}\{ \boldsymbol{\tau}^H\mathbf{w} \} - \frac{\rho}{2}\mathbf{w}^H\mathbf{w}.\nonumber
\end{align}

The optimal solution for (P7) is given by the following Theorem \ref{theorem_1}, 
which is proved in Appendix A.

\begin{theorem} \label{theorem_1}
 If $- 2\text{Re}\{ \mathbf{b}_{4}^H (\bar{\mathbf{B}}_{9}^{-1}\bar{\mathbf{b}}_{3}) \}- c_{3} \leq 0    $,
 the optimal solution is written as
 \begin{align}
\mathbf{f}_{IE}^{\star} 
=
\bar{\mathbf{B}}_{9}^{-1}\bar{\mathbf{b}}_{3},
\end{align}
otherwise, 
we can have
 \begin{align}
\mathbf{f}_{IE}^{\star} 
=
\bar{\mathbf{B}}_{9}^{-1}( \kappa \mathbf{b}_{4} + \bar{\mathbf{b}}_{3}),
\end{align}
where
the value of $\kappa$  can be obtained by the Newton's method.
\end{theorem}

With 
$\mathbf{f}_{IE}$
and 
$\boldsymbol{\tau}$
being given,
the optimization problem w.r.t. $\mathbf{w}$ can be represented as
\begin{subequations}
\begin{align}
\textrm{(P8)}:&\mathop{\textrm{min}}
\limits_{ \mathbf{w}
}\
\frac{\rho}{2}  \mathbf{w}^H \mathbf{w}
-
 2\text{Re}\{ {\mathbf{b}}_{5}^H \mathbf{w} \} - {c}_{4}\label{P8_obj}
\\
\textrm{s.t.}\ 
& \mathbf{w}^H\mathbf{\bar{A}}_{IE,n}\mathbf{w} \leq P_t, \forall n \in \mathcal{N},
\end{align}
\end{subequations}
where
\begin{align}
& {c}_{4}
\triangleq 
\mathbf{f}_{IE}^H \mathbf{B}_{9} \mathbf{f}_{IE}
-
 2\text{Re}\{ \mathbf{b}_{3}^H \mathbf{f}_{IE} \} - c_{2}\\
& + \text{Re}\{ \boldsymbol{\tau}^H\mathbf{f}_{IE}   \}
 + \frac{\rho}{2} \Vert\mathbf{f}_{IE}  \Vert_2^2,
{\mathbf{b}}_{5}
\triangleq 
\frac{1}{2}
(\boldsymbol{\tau} +\rho \mathbf{f}_{IE} ).\nonumber
\end{align}

Obviously, 
the optimization problem (P8) can be decomposed into $N$ independent subproblems,
which can be solved independently in parallel.
As such,
each subproblem can be formulated as
\begin{subequations}
\begin{align}
\textrm{(P9)}:&\mathop{\textrm{min}}
\limits_{ \mathbf{w}_n
}\
\mathbf{w}_n^H \mathbf{w}_n
-
 2\text{Re}\{ {\mathbf{b}}_{6,n}^H \mathbf{w}_n \}\label{P9_obj}
\\
\textrm{s.t.}\ 
& \mathbf{w}_n^H\mathbf{w}_n \leq P_t,\label{P9_c_1}
\end{align}
\end{subequations}
where
\begin{align}
& \mathbf{b}_{6} 
\triangleq 
\frac{2}{\rho}
\mathbf{b}_{5},\\
& \mathbf{w}_n
\triangleq 
\big[ \mathbf{w}(n),\mathbf{w}(n+N),\cdots, \mathbf{w}(n+(K-1)N), \nonumber\\
&\cdots, \mathbf{w}(n+(K+G-1)N)  \big]^T \in \mathbb{C}^{(K+G)\times 1},\nonumber\\
& \mathbf{b}_{6,n}
\triangleq 
\big[ \mathbf{b}_{6}(n),\mathbf{b}_{6}(n+N),\cdots, \mathbf{b}_{6}(n+(K-1)N),\nonumber \\
&\cdots, \mathbf{b}_{6}(n+(K+G-1)N)  \big]^T \in \mathbb{C}^{(K+G)\times 1}.\nonumber
\end{align}

The analytical solution of (P9) can be directly obtained by the following theorem
that is proved in Appendix B.

\begin{theorem} \label{theorem_2}
If the inequality $ \mathbf{b}_{6,n}^H\mathbf{b}_{6,n} \leq P_t $  is satisfied,
the optimal solution of (P9) is given as
\begin{align}
\mathbf{w}_n^{\star}
=
\mathbf{b}_{6,n}.
\end{align}

Otherwise,
the optimal solution to problem (P9) can be directly expressed as 
\begin{align}
\mathbf{w}_n^{\star}
=
\sqrt{P_t} 
\frac{\mathbf{b}_{6,n}}{\Vert\mathbf{b}_{6,n} \Vert}. \label{ADMM_rho}
\end{align}

\end{theorem}

As proposed in \cite{ref_ADMM} for the ADMM method, 
after the primal variables being updated,
the dual variable $\boldsymbol{\tau}$ can be updated via the gradient ascent method,
which is formulated as
\begin{align}
\boldsymbol{\tau}^{t+1} := \boldsymbol{\tau}^{t} + \rho(\mathbf{f}_{IE}-\mathbf{w}).
\end{align}

The ADMM-based low complexity solution for solving problem (P3) 
is summarized in Algorithm \ref{alg:3}.

\begin{algorithm}[t]
\caption{The ADMM-based Method}
\label{alg:3}
\begin{algorithmic}[1]
\STATE {initialize}
$\mathbf{f}_{IE}^{(0)}$,
$\mathbf{w}^{(0)}$,
$\boldsymbol{\tau}^{(0)}$
and
$t=0$
;
\REPEAT
\STATE update $\mathbf{f}_{IE}^{(t)}$ by solving (P7);
\FOR{ $n = 1:N$ }
\STATE update $\mathbf{w}_{n}^{(t)}$ by solving (P9);
\ENDFOR
\STATE update $\boldsymbol{\tau}^{(t)}$ by  (\ref{ADMM_rho});
\STATE $t++$;
\UNTIL{$convergence$;}
\end{algorithmic}
\end{algorithm}

\section{Numerical Results}

\begin{figure}[t]
	\centering
	\includegraphics[width=.49\textwidth]{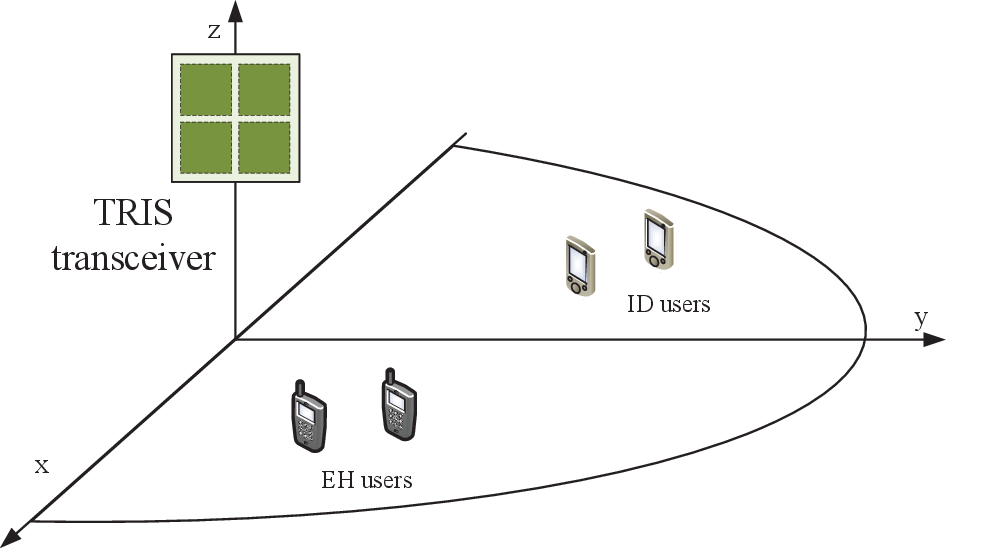}
	\caption{Simulation setup for a TRIS transceiver-enabled SWIPT system.}
	\label{fig.2}
\end{figure}


In this section, 
extensive numerical results will be provided to verify the effectiveness of our proposed algorithms
for the considered TRIS transceiver-enabled SWIPT system.
The simulated system setting of the TRIS transceiver-enabled SWIPT system is shown in Fig. \ref{fig.2},
which includes one TRIS transceiver, $K=2$ ID users and $G=2$ EH users.
Due to the higher power requirement of EH users compared to ID users, 
EH users are deployed near the TRIS transceiver.
In the experiment,
the TRIS transceiver is located at the three-dimensional (3D) coordinates $(0,0,4.5)$m. 
All ID users are randomly distributed in a sector spanning distances from $20$m to $50$m and are placed at a height of $1.5$ m.
The antenna spacing is set to half the wavelength of the carrier.
The path loss exponent of the TRIS transceiver-ID user and the TRIS transceiver-EH user channels are set as $3.2$ and $2.2$, 
respectively. 
The transmit power limit for each antenna of the TRIS transceiver is set as $10$dBm. 
The noise level at the ID users is set as $-90$dBm.

\begin{figure}[t]
	\centering
	\includegraphics[width=.52\textwidth]{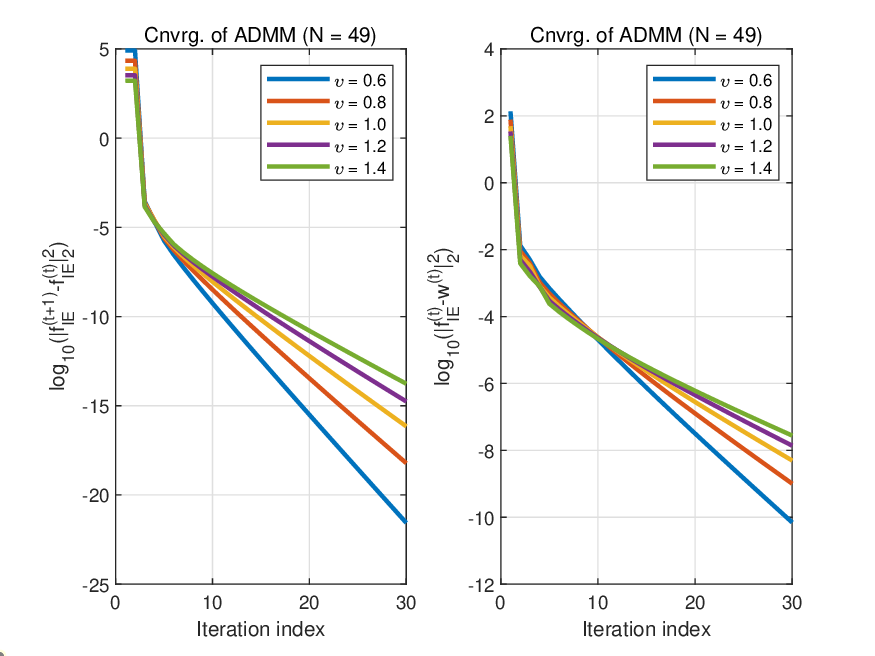}
	\caption{Convergence of ADMM Algorithm.}
	\label{fig.3}
\end{figure}

\begin{figure}[t]
	\centering
	\includegraphics[width=.52\textwidth]{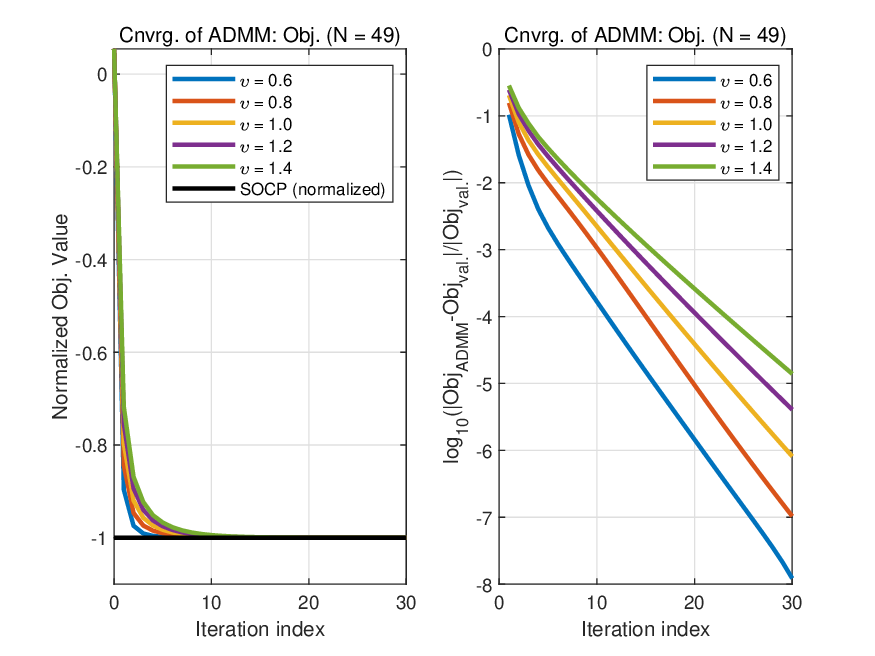}
	\caption{Convergence of objective value of ADMM method.}
	\label{fig.4}
\end{figure}

First,
both Fig. \ref{fig.3} and Fig. \ref{fig.4} jointly present 
the convergence behavior of the analytical solution Alg. \ref{alg:3} employed to update the beamforming 
by leveraging the ADMM methodology.
For assuring the fair comparison, 
both the SOCP and low-complexity implementations start from one common initial point in each channel realization.
In Fig. \ref{fig.3}, 
under different values of coefficient $\rho$,
the left and right subfigures illustrate the difference 
$\vert \mathbf{f}_{IE}^{(t+1)}-\mathbf{f}_{IE}^{(t)} \vert_2^2$
and
$\vert \mathbf{f}_{IE}^{(t)}-\mathbf{w}^{(t)} \vert_2^2$ in the log domain, 
respectively,
as ADMM iterations progress. 
These plots illustrate how both the change in the iterative beamforming vector 
and the primal-dual residual between $\mathbf{f}_{IE}^{(t)}$ and  $\mathbf{w}^{(t)}$
evolve as ADMM iterations proceed, 
thereby providing insight into the effect of $\rho$ on convergence speed and stability.
By comparing these curves for different $\rho$ values, 
one can assess how  parameter $\rho$ influences convergence rate, 
residual magnitude, 
and numerical stability of the ADMM-based algorithm.
As shown in Fig. \ref{fig.3}, 
an appropriate value 
lies in the range of [0.6, 1.4], 
which yields sufficient convergence 
(e.g., to a precision of $10^{-6}$) within several tens of iterations.

Fig. \ref{fig.4} provides a detailed view of the objective convergence behavior yielded by Alg. \ref{alg:3}.
The left subfigure plots the evolution of the objective values for several values of the coefficient $\rho$.
Note the black line represents the true objective value of (P4) by solving the SOCP  (implemented using CVX).
This value is used as a benchmark and has been normalized for comparison.
The right subfigure displays the difference between the true objective value produced by Alg. \ref{alg:3} in the log domain, 
making it easy to observe the change in the residual over iterations.
In general,
as shown in Fig. \ref{fig.4},
Alg. \ref{alg:3} yields sufficiently accurate objective value rapidly.
Specifically, 
the algorithm, across the tested values of $\rho$,
attains a highly accurate objective value within roughly 15 iterations.
The log-scale residual plot further 
reveals that the error decays by several orders of magnitude during the first few iterations and then tapers off, 
indicating fast initial progress followed by steady refinement toward the SOCP benchmark. 
These results demonstrate both the robustness of the method to the choice of $\rho$ 
and its practical efficiency for solving (P4).

\begin{figure}[t]
	\centering
	\includegraphics[width=.52\textwidth]{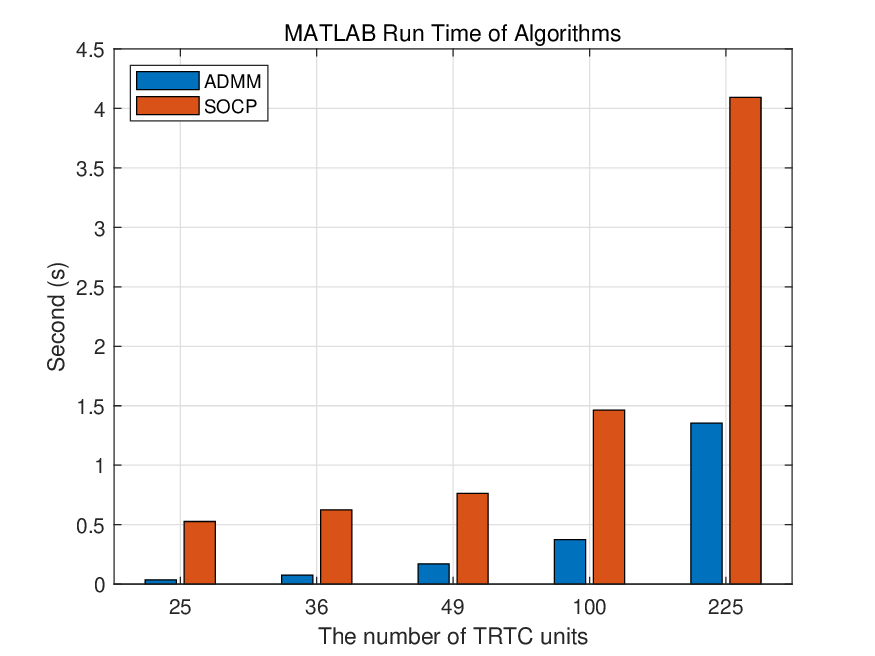}
	\caption{Comparison of MATLAB run time for proposed algorithms.}
	\label{fig.5}
\end{figure}

In Fig. \ref{fig.5},
we examine the computational complexity of the analytic-based solution Alg. \ref{alg:3}.
Fig. \ref{fig.5} presents the MATLAB execution times for both CVX and Alg. \ref{alg:3} 
under different values of the number of TRIS transceiver architecture elements $N$.
As reflected by the results,
across the tested configurations,
both implementations' runtimes increase.
However,
Alg. \ref{alg:3} is substantially faster:
its runtime is smaller by roughly one to two orders of magnitude compared with the SOCP solver.
This speed-up highlights the practical advantage of the analytic approach for larger-scale TRIS transceiver configurations.

\begin{figure}[t]
	\centering
	\includegraphics[width=.52\textwidth]{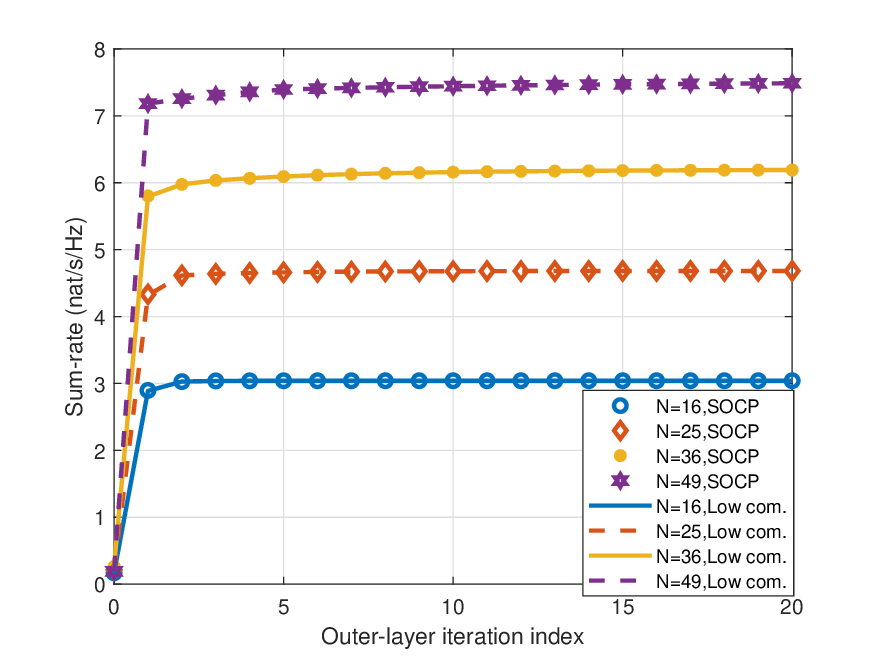}
	\caption{Convergence of Algorithms.}
	\label{fig.6}
\end{figure}

Fig. \ref{fig.6} examines the overall convergence behavior of the proposed algorithms 
for solving the original problem (P0), 
comparing the SOCP-based solution with the low-complexity (Low com.) analytic solution across several TRIS transceiver unit sizes.
For a fair comparison, 
both methods are initialized from identical starting point in every test.
As seen from the figure,
the SOCP-based and analytic-based solutions produce virtually indistinguishable performance for each element size,
indicating that the analytic approximation preserves the solution quality of the SOCP formulation.
Furthermore,
both algorithms converge rapidly
that the sum-rate reaches a steady value within roughly 10 outer-layer iterations.
In particular, 
small configurations (e.g., $N=16$) converge in about 3-5 iterations, 
whereas larger configurations 
(e.g., $N=49$) display a slightly steeper initial rise but still saturate well before the 10-th iteration.
These results demonstrate two important points. 
On the one hand, 
the low-complexity analytic method reliably attains near-optimal sum-rate performance 
comparable to the SOCP-based approach; 
on the other hand, 
the fast convergence exhibited by both methods implies low iterative overhead in practice. 
Combined with the substantially lower per-iteration runtime of the analytic algorithm (see Fig. \ref{fig.5}), 
the close match in convergence behavior makes 
the Low-com. solution particularly attractive for real-time or large-scale TRIS transceiver deployments 
where computational resources are constrained.

\begin{figure}[t]
	\centering
	\includegraphics[width=.52\textwidth]{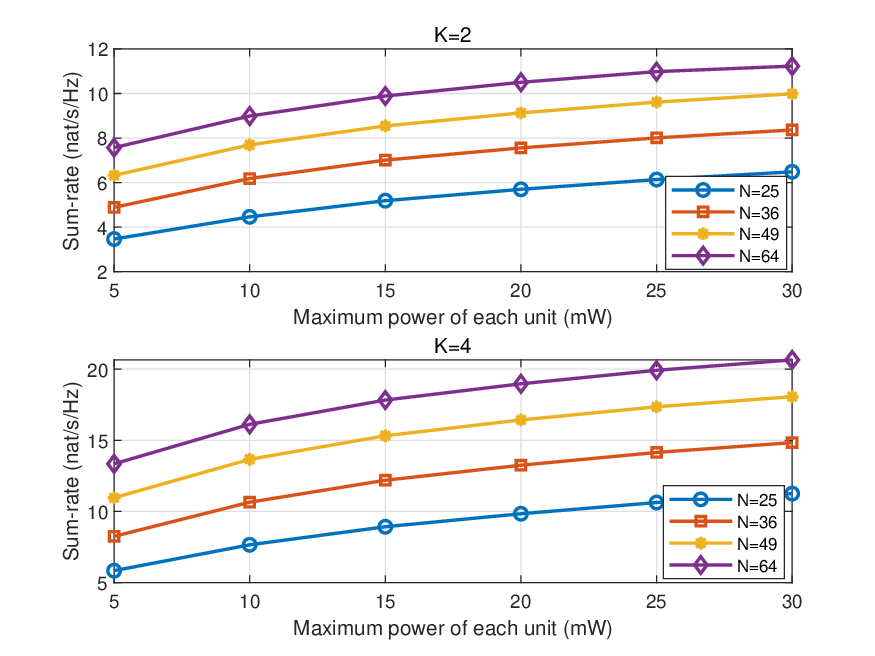}
	\caption{Sum-rate versus the maximum power of each TRIS unit.}
	\label{fig.7}
\end{figure}

Fig. \ref{fig.7} explores 
the relationship between the maximum available transmit power of 
the TRIS unit and the corresponding sum-rate.
The upper and lower subplots correspond to different numbers of ID users, respectively.
In our test, 
the maximum available transmit power of the TRIS unit varies from 5 mW to 30 mW.
As shown in Fig. \ref{fig.7},
it can be observed that,
for every considered array size,
the sum-rate monotonically increases as the TRIS unit's maximum transmit power increases.
This behavior stems from the increased maximum transmit power of the TRIS unit, 
which enables the system to provide greater diversity gain to ID users and thus leads to a higher sum-rate performance.
The incremental gain from increasing transmit power diminishes at higher power values, 
as evidenced by the flattening of the curves near 30 mW, 
which implies that performance becomes progressively limited 
by interference and channel conditions rather than transmit power alone.
Moreover, 
for any fixed power level, 
larger arrays consistently yield higher sum-rate, 
indicating additional beamforming and diversity gains afforded by increased aperture.

\begin{figure}[t]
	\centering
	\includegraphics[width=.52\textwidth]{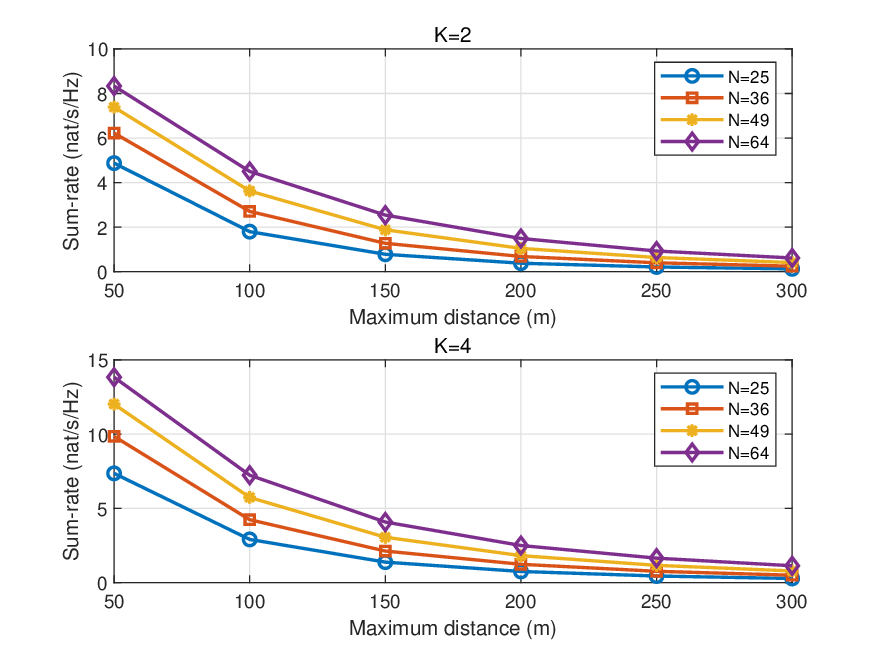}
	\caption{Sum-rate versus the maximum distance between TRIS and ID users.}
	\label{fig.8}
\end{figure}

In Fig. \ref{fig.8}, 
we illustrate the sum-rate performance versus the maximum distance between TRIS and ID users 
for two ID user configurations (i.e., $K=2$ and $K=4$).
It is observed that the sum-rate decreases monotonically with increasing distance from 50m to 300m
for all array sizes $N$, with the most pronounced decline occurring in the 50-100 m interval.
At any fixed distance, 
larger arrays consistently yield higher sum-rate performance, 
indicating that increased aperture provides additional beamforming 
and diversity gains that partially compensate for propagation loss.
For distances beyond approximately 150-200 m, 
the curves for different $N$ converge toward low sum-rate values, 
indicating diminishing returns from increasing array size as the system becomes power- or channel-limited.
These results underscore that while enlarging the TRIS aperture mitigates distance-induced performance loss, 
proximity between TRIS and ID users remains a primary determinant of achievable throughput.

\begin{figure}[t]
	\centering
	\includegraphics[width=.52\textwidth]{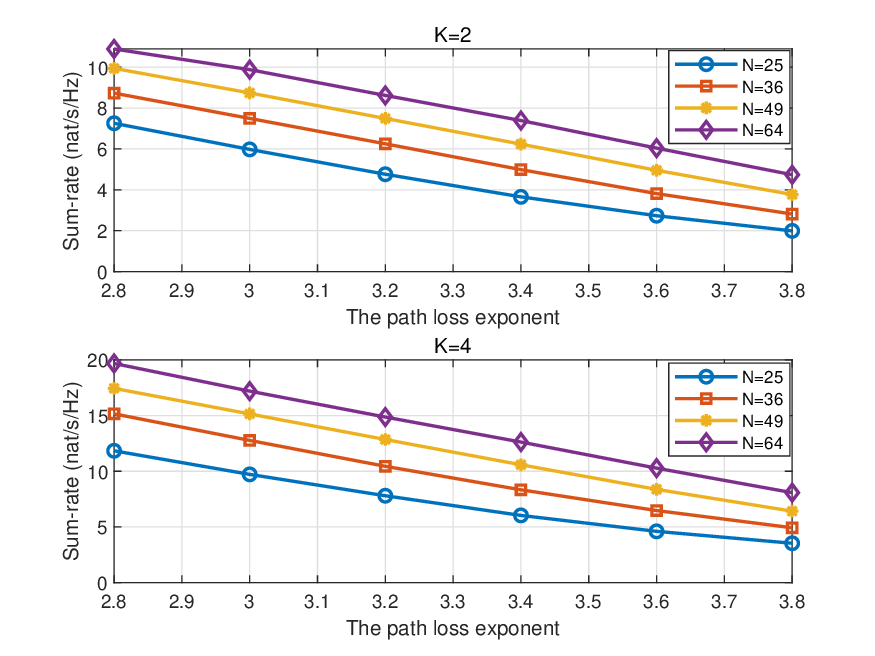}
	\caption{Sum-rate versus the path loss exponent.}
	\label{fig.9}
\end{figure}

Fig. \ref{fig.9} investigates the impact of the path loss exponent of the TRIS transceiver-ID user channel on the sum-rate 
under different system configurations.
Specifically, 
the Fig. \ref{fig.9} presents 
the sum-rate as a function  of the path loss exponent varying from 2.8 to 3.8, 
for two different numbers of ID users $K=2$ and $K=4$,
and multiple antenna array sizes $N=25, 36, 49, 64$.
As observed, 
the sum-rate demonstrates a clear monotonic decrease with the increasing path loss exponent. 
This behavior is attributable to the fact that a larger path loss exponent 
corresponds to more severe signal attenuation over distance, 
thereby reducing the effective received signal power at the ID users. 
Moreover, 
it can be seen that for a fixed number of users, increasing the antenna array size $N$
consistently improves the sum-rate performance, 
which is expected due to the enhanced spatial diversity and beamforming gain. 
However, 
the negative impact of a higher path loss exponent remains evident regardless of the system scale.

\section{Conclusions}

In this paper,
we investigate an innovative TRIS transceiver-enabled SWIPT system with multiple ID and EH users.
Specifically,
the goal of this paper is to maximize the sum-rate of all ID users 
by designing the transmit beamforming vectors produced by the TRIS transceiver architecture,
subject to 
the minimum energy harvesting constraint of all EH users
and the per-antenna power constraints of each TRIS transceiver unit.
Per-element power constraints are imposed to accurately model the limited amplification capability of each TRIS transceiver element, 
significantly increasing the number of constraints and rendering the optimization problem highly challenging. 
To address the non-convex sum-rate optimization problem,
we firstly equivalently transform the rate function via the WMMSE method, 
and then alternately update the original and auxiliary variables combined with the MM methodology.
To reduce the computation complexity,
we successfully develop an analytic-based and highly computation-parallelizable algorithm based on the ADMM framework.
Numerical results demonstrated that our proposed optimization algorithms can significantly improve sum-rate performance.
Moreover,
the ADMM-based algorithm can achieve reduced complexity without
performance loss  compared to the SOCP-based method.

\appendix
\subsection{Proof of Theorem 1}
\normalem
Proof:
Consider the following optimization problem
\begin{subequations}
\begin{align}
(\textrm{P}_{Lm1}):\mathop{\textrm{min}}
\limits_{\mathbf{x}}\
& \mathbf{x}^H\mathbf{D}\mathbf{x}-2\textrm{Re}\{\mathbf{d}^H\mathbf{x}\}+d\\
\textrm{s.t.}\
&\mathbf{x}^H\mathbf{\bar{D}}\mathbf{x}-2\textrm{Re}\{\mathbf{\bar{d}}^H\mathbf{x}\}+\bar{d}\leq 0,
\end{align}
\end{subequations}
where the matrix $\mathbf{D}$ is strictly positive definite (i.e., $\mathbf{D} \succ {0}$),
and the matrix $\mathbf{\bar{D}}$ is positive semidefinite (i.e., $\mathbf{\bar{D}}\succcurlyeq {0}$).

Since the Slater's condition of problem ($\textrm{P}_{Lm1}$) is satisfied, 
and then strong duality holds. 
Consequently, 
the Karush-Kuhn-Tucker (KKT) conditions' analysis \cite{ref_Convex Optimization} 
can be employed to find the optimal solution of the optimization problem ($\textrm{P}_{Lm1}$).
First, 
the Lagrange function associated with the problem ($\textrm{P}_{Lm1}$) is given as
\begin{align}
\mathcal{L}( \mathbf{x}, \nu) =& 
\mathbf{x}^H\mathbf{D}\mathbf{x}-2\textrm{Re}\{\mathbf{d}^H\mathbf{x}\}+d\\
&+ \nu( \mathbf{x}^H\mathbf{\bar{D}}\mathbf{x}-2\textrm{Re}\{\mathbf{\bar{d}}^H\mathbf{x}\}+\bar{d}  ), \nonumber
\end{align}
where 
$\nu$ denotes the Lagrangian multiplier.
Then the KKT conditions of the optimization problem  ($\textrm{P}_{Lm1}$) are formulated as follows
\begin{subequations} \label{Proof_Lemmma1_KKT}
\begin{align}
& \nu \geq 0,
\mathbf{x}^H\mathbf{\bar{D}}\mathbf{x}-2\textrm{Re}\{\mathbf{\bar{d}}^H\mathbf{x}\}+\bar{d}\leq 0,\label{Proof_Lemma1_0}\\
& \nu ( \mathbf{x}^H\mathbf{\bar{D}}\mathbf{x}-2\textrm{Re}\{\mathbf{\bar{d}}^H\mathbf{x}\}+\bar{d} ) = 0,\label{Proof_Lemma1_1}\\
& \mathbf{{D}}\mathbf{x}-\mathbf{{d}}
+\nu ( \mathbf{\bar{D}}\mathbf{x}-\mathbf{\bar{d}} ) = 0.\label{Proof_Lemma1_2}
\end{align}
\end{subequations}

Since the Lagrangian multiplier $\varsigma$ is non-negative,
we can analyze the KKT conditions in two possible cases according to the sign value of $\varsigma$,
i.e.,
$\varsigma = 0$
or 
$\varsigma >0$.

\begin{itemize}
\item[] {CASE-I}:
When $\varsigma^{\star}=0$,
by leveraging the equation (\ref{Proof_Lemma1_2}),
we have 
\begin{align}
\mathbf{x}^{\star}=\mathbf{D}^{-1}\mathbf{d}.\label{Proof_Lemma1_3}
\end{align}

At this time,
the KKT conditions (\ref{Proof_Lemmma1_KKT}) hold
if and only if the inequality 
$(\mathbf{x}^{\star})^{H}\mathbf{\bar{D}}\mathbf{x}^{\star}
-2\textrm{Re}\{\mathbf{\bar{d}}^H\mathbf{x}^{\star}\}+\bar{d}\leq 0$ 
is satisfied.
If the inequality is satisfied, 
then the candidate solution $\mathbf{x}^{\star}$ given in (\ref{Proof_Lemma1_3}) is indeed optimal. 
Conversely, 
if the KKT conditions  (\ref{Proof_Lemmma1_KKT}) cannot be fulfilled, 
we proceed to analyze the case $\varsigma^{\star}>0$.

\item[] {CASE-II}: 
When $\varsigma^{\star}>0$, by using (\ref{Proof_Lemma1_1}) and (\ref{Proof_Lemma1_2}),
we can obtain
\begin{align}
\mathbf{x}^{\star} = (\varsigma^{\star}\mathbf{\bar{D}}+\mathbf{D})^{-1}(\varsigma^{\star}\mathbf{\bar{d}}+\mathbf{d}).\label{Proof_Lemma1_4}
\end{align}

Since the following inequality will be hold:
\begin{align}
& (\mathbf{D}^{-1}\mathbf{{d}})^{ H}\mathbf{\bar{D}}(\mathbf{{D}}^{-1}\mathbf{{d}})\\
&-2\textrm{Re}\{\mathbf{\bar{d}}^H(\mathbf{{D}}^{-1}\mathbf{{d}})\}+\bar{q}> 0, \nonumber
\end{align}
hence there have a unique positive $\varsigma^{\star}$ to satisfy the following equality
\begin{align}
&\big((\varsigma^{\star}\mathbf{\bar{D}}+\mathbf{D})^{-1}(\varsigma^{\star}\mathbf{\bar{d}}+\mathbf{d})\big)^H
\mathbf{\bar{D}}
(\varsigma^{\star}\mathbf{\bar{D}}+\mathbf{D})^{-1}(\varsigma^{\star}\mathbf{\bar{d}}+\mathbf{d})\nonumber\\
&-2\textrm{Re}\{\mathbf{\bar{d}}^H(\varsigma^{\star}\mathbf{\bar{D}}+\mathbf{D})^{-1}(\mathbf{\varsigma^{\star}\bar{d}}+\mathbf{d})\}+\bar{d}= 0.
\end{align}
The unique $\varsigma^{\star}$ can be efficiently obtained by the  Newton's method.
\end{itemize}

Thus, the proof of  Lemma \ref{theorem_1} is completed.

\subsection{Proof of Theorem 2}
\normalem
Proof:
First,
the Lagrange function associated with the problem (P9) can be formulated as
\begin{align}
\mathcal{L}(\mathbf{w}_{n},\mu) =& \mathbf{w}_n^H \mathbf{w}_n
-
 2\text{Re}\{ {\mathbf{b}}_{6,n}^H \mathbf{w}_n \}\\
 &+ \mu (\mathbf{w}_n^H\mathbf{w}_n - P_t)
,\nonumber 
\end{align}
where $\mu$ is the Lagrangian multiplier.

Next,
by setting the first-order derivative of the Lagrange function $\mathcal{L}(\mathbf{w}_{n},\mu)$ 
w.r.t. the variable $\mathbf{w}_{n}$ to zero,
we can obtain the following formulation
\begin{align}
\frac{\partial \mathcal{L}(\mathbf{w}_{n},\mu)}{\partial \mathbf{w}_{n} } = \mathbf{0}.
\end{align}

Furthermore,
the solution of $\mathbf{w}_{n}$ can be represented as follows
\begin{align}
\mathbf{w}_{n} = \frac{\mathbf{b}_{6,n}}{ 1 + \mu  }. \label{P14_closed_solution}
\end{align}

By substituting the optimal solution  (\ref{P14_closed_solution}) into the power constraint (\ref{P9_c_1}), 
we can have the following expression
\begin{align}
\frac{\mathbf{b}_{6,n}^H\mathbf{b}_{6,n}}{ (1+\mu )^2  }\leq P_t.  \label{P14_proof_power}
\end{align}

Following (\ref{P14_proof_power}), 
we can observe that
the left hand side of  (\ref{P14_proof_power}) is a decreasing function w.r.t. 
the Lagrangian multiplier $\mu$.
Then
the optimal solution to problem (P9) is given by one of the
following two cases:
\begin{itemize}
\item[] \underline{CASE-I}:
if $\mu = 0$,
the inequality (\ref{P14_proof_power}) is satisfied.
The optimal solution of (P9) is given as
\begin{align}
\mathbf{w}_{n}^{\star} = \mathbf{b}_{6,n}. 
\end{align}

\item[] \underline{CASE-II}: 
Otherwise,
$\mu $ is positive.
And the optimal solution of problem (P9) is
\begin{align}
\mathbf{w}_{n}^{\star} = \sqrt{P_t}\frac{\mathbf{b}_{6,n}}{  \Vert\mathbf{b}_{6,n}\Vert_2  }. 
\end{align}

\end{itemize}



\begin{thebibliography}{99}



\bibitem{ref_SWIPT_1}
I. Krikidis, S. Timotheou, S. Nikolaou, G. Zheng, D. W. K. Ng, and R. Schober, 
``Simultaneous wireless information and power transfer in modern communication systems,'' 
\emph{IEEE Commun. Mag.}, 
vol. 52, no. 11, pp. 104$-$110, Nov. 2014.



\bibitem{ref_SWIPT_2}
R. Zhang and C. K. Ho, 
``MIMO broadcasting for simultaneous wireless information and power transfer,'' 
\emph{IEEE Trans. Wireless Commun.}, 
vol. 12, no. 5, pp. 1989$-$2001, May 2013.


\bibitem{ref_RIS_1}
Q. Wu \emph{et al.}, 
``Intelligent surfaces empowered wireless network: Recent advances and the road to 6G,''
\emph{Pro. IEEE}, 
vol. 112, no. 7, pp. 724$-$763, Jul. 2024.

\bibitem{ref_RIS_2}
Q. Wu, S. Zhang, B. Zheng, C. You, and R. Zhang, 
``Intelligent reflecting surface-aided wireless communications: A tutorial,'' 
\emph{IEEE Trans. Commun.}, 
vol. 69, no. 5, pp. 3313$-$3351, May 2021.




\bibitem{ref_RIS_app_1}
G. Zhou, C. Pan, H. Ren, K. Wang, and A. Nallanathan, 
``Intelligent reflecting surface aided multigroup multicast MISO communication systems,'' 
\emph{IEEE Trans. Signal Process.}, 
vol. 68, pp. 3236$-$3251, Apr. 2020.


\bibitem{ref_RIS_app_2}
S. Gong, C. Xing, P. Yue, L. Zhao, and T. Q. S. Quek, 
``Hybrid analog and digital beamforming for RIS-assisted mmWave communications,''
\emph{IEEE Trans. Wireless Commun.}, 
vol. 22, no. 3, pp. 1537$-$1554, Mar. 2023.

\bibitem{ref_RIS_app_3}
Y. Guo, Y. Liu, Q. Wu, Q. Shi, and Y. Zhao, 
``Enhanced secure communication via novel double-faced active RIS,'' 
\emph{IEEE Trans. Commun.}, 
vol. 71, no. 6, pp. 3497$-$3512, Jun. 2023.

\bibitem{ref_RIS_app_4}
J. Wang, J. Xiao, Y. Zou, W. Xie, and Y. Liu, 
``Wideband beamforming for RIS assisted near-field communications,'' 
\emph{IEEE Trans. Wireless Commun.}, 
vol. 23, no. 11, pp. 16836$-$16851, Nov. 2024.

\bibitem{ref_RIS_app_5}
Y. Guo, Y. Liu, Q. Wu, X. Li, and Q. Shi, 
``Joint beamforming and power allocation for RIS aided full-duplex integrated sensing and uplink communication system,'' 
\emph{IEEE Trans. Wireless Commun.}, 
vol. 23, no. 5, pp. 4627$-$4642, May 2024.

\bibitem{ref_RIS_app_6}
Z. Zhang, W. Chen, Q. Wu, Z. Li, X. Zhu, and J. Yuan, 
``Intelligent omni surfaces assisted integrated multi-target sensing and multi-user MIMO communications,'' 
\emph{IEEE Trans. Commun.}, 
vol. 72, no. 8, pp. 4591$-$4606, Aug. 2024.

\bibitem{ref_RIS_app_7}
Z. Guang, Y. Liu, Q. Wu, Y. -F. Liu, and Q. Shi, 
``Communication aided sensing for RIS assisted MU-MIMO system: CRB optimization with guaranteed ergodic rate,'' 
\emph{IEEE Trans. Signal Process.}, 
early access,
September 29, 2025,
doi: 10.1109/TSP.2025.3614347.

\bibitem{ref_RIS_SWIPT_1}
Q. Wu and R. Zhang, 
``Weighted sum power maximization for intelligent reflecting surface aided SWIPT,'' 
\emph{IEEE Wireless Commun. Lett.}, 
vol. 9, no. 5, pp. 586$-$590, May 2020.



\bibitem{ref_RIS_SWIPT_2}
Q. Wu and R. Zhang, 
``Joint active and passive beamforming optimization for intelligent reflecting surface assisted SWIPT under QoS constraints,'' 
\emph{IEEE J. Sel. Areas  Commun.}, 
vol. 38, no. 8, pp. 1735$-$1748, Aug. 2020.

\bibitem{ref_RIS_SWIPT_3}
C. Pan \emph{et al.}, 
``Intelligent reflecting surface aided MIMO broadcasting for simultaneous wireless information and power transfer,'' 
\emph{IEEE J. Sel. Areas  Commun.}, 
vol. 38, no. 8, pp. 1719$-$1734, Aug. 2020.


\bibitem{ref_RIS_SWIPT_4}
Z. Li, W. Chen, Q. Wu, K. Wang, and J. Li, 
``Joint beamforming design and power splitting optimization in IRS-assisted SWIPT NOMA networks,'' 
\emph{IEEE Trans. Wireless Commun.}, 
vol. 21, no. 3, pp. 2019$-$2033, Mar. 2022.

\bibitem{ref_RIS_SWIPT_5}
Y. Guo, Y. Liu, M. Li, Q. Wu, and Q. Shi, 
``Beamforming design for power transferring and secure communication in RIS-aided network,'' 
in \emph{Proc. IEEE Int. Conf. Commun. (ICC)}, 
Seoul, Korea, Republic of, 2022, pp. 450$-$455.

\bibitem{ref_RIS_SWIPT_6}
Y. Li, J. Wang, Y. Zou, W. Xie, and Y. Liu, 
``Weighted sum power maximization for STAR-RIS assisted SWIPT systems,'' 
\emph{IEEE Trans. Wireless Commun.}, 
vol. 23, no. 12, pp. 18394$-$18408, Dec. 2024.


\bibitem{ref_RIS_SWIPT_7}
C. Huang, W. Chen, Q. Wu, X. Zhu, Z. Li, Y. Wang, and J. Yuan,
``Dual-IRS aided near-/hybrid-field SWIPT: Passive beamforming and independent antenna power splitting design,''
Aug. 2025.
[Online]. 
Available: https://arxiv.org/abs/2508.20531


\bibitem{ref_TRIS_1}
Z. Li \emph{et al.}, 
``Transmissive reconfigurable intelligent surface-enabled transceiver systems: Architecture, design issues, and opportunities,''
\emph{IEEE Veh. Technol. Mag.}, 
vol. 19, no. 4, pp. 44$-$53, Dec. 2024.

\bibitem{ref_TRIS_2}
X. Bai, F. Kong, Y. Sun, G. Wang, J. Qian, X. Li, A. Cao, C. He, X. Liang, R. Jin, and W. Zhu, 
``High-efficiency transmissive programmable metasurface for multimode OAM generation,'' 
\emph{Adv. Opt. Mater.}, 
vol. 8, no. 17, p. 2000570, Jun. 2020.

\bibitem{ref_TRIS_3}
X. Bai, F. Zhang, L. Sun, A. Cao, J. Zhang, C. He, L. Liu, J. Yao, and W. Zhu, 
``Time-modulated transmissive programmable metasurface for low sidelobe beam scanning,'' 
\emph{Research}, 
Jul. 2022.

\bibitem{ref_TRIS_app_1}
Z. Li \emph{et al.}, 
``Toward TMA-based transmissive RIS transceiver enabled downlink communication networks: A consensus-ADMM approach,''
\emph{IEEE Trans. Commun.}, 
vol. 73, no. 4, pp. 2832$-$2846, Apr. 2025.

\bibitem{ref_TRIS_app_2}
Z. Li, W. Chen, Z. Zhang, Q. Wu, H. Cao, and J. Li, 
``Robust sum-rate maximization in transmissive RMS transceiver-enabled SWIPT networks,'' 
\emph{IEEE Internet Things J.}, 
vol. 10, no. 8, pp. 7259$-$7271, Apr. 2023.

\bibitem{ref_TRIS_app_3}
Z. Li, W. Chen, Z. Liu, H. Tang, and J. Lu, 
``Joint communication and computation design in transmissive RMS transceiver enabled multi-tier computing networks,'' 
\emph{IEEE J. Sel. Areas Commun.}, 
vol. 41, no. 2, pp. 334$-$348, Feb. 2023.

\bibitem{ref_TRIS_app_4}
A. Huang, X. Mu, L. Guo, and G. Zhu, 
``Hybrid active-passive RIS transmitter enabled energy-efficient multi-user communications,'' 
\emph{IEEE Trans. Wireless Commun.}, 
vol. 23, no. 9, pp. 10653$-$10666, Sep. 2024.

\bibitem{ref_TRIS_app_5}
Y. Wang, S. Yang, Z. Chu, B. Ji, M. Hua, and C. Li, 
``Robust weighted sum secrecy rate maximization for joint ITS- and IRS-assisted multiantenna networks,'' 
\emph{IEEE Wireless Commun. Lett.}, 
vol. 14, no. 3, pp. 681$-$685, Mar. 2025.

\bibitem{ref_TRIS_app_6}
Z. Liu  \emph{et al.}, 
``Enhancing robustness and security in ISAC network design: Leveraging transmissive reconfigurable intelligent surface with
RSMA,'' 
\emph{IEEE Trans. Commun.}, 
early access, 
March 31, 2025, 
doi: 10.1109/TCOMM.2025.3555894.

\bibitem{ref_TRIS_app_7}
Z. Liu, W. Chen, Q. Wu, Z. Li, Q. Wu, N. Cheng, and J. Li,
``Beamforming design and multi-user scheduling in transmissive RIS
enabled distributed cooperative ISAC networks with RSMA,'' 
Nov. 2024.
[Online]. Available: https://arxiv.org/abs/2411.10960

\bibitem{ref_TRIS_app_8}
M. Asif, X. Bao, Z. Ali, A. Ihsan, M. Ahmed, and X. Li, 
``Transmissive RIS-empowered LEO-satellite communications with hybrid-NOMA under residual hardware impairments,'' 
\emph{IEEE Trans. Green Commun. Netw.},
early access, 
September 23, 2024, 
doi: 10.1109/TGCN.2024.3466469.

\bibitem{ref_TRIS_app_9}
X. Zhu, Q. Wu, and W. Chen, 
``Transmissive RIS transmitter enabled spatial modulation MIMO systems,'' 
\emph{IEEE J. Sel. Areas Commun.}, 
vol. 43, no. 3, pp. 899$-$911, Mar. 2025.


\bibitem{ref_TRIS_app_10}
Y. Guo, W. Chen, Q. Wu, Y. Zhu, Y. Liu, Z. Li, and Y. Wang,
``Max-min rate optimization for multigroup multicast MISO systems via novel transmissive RIS transceiver,'' 
Jul. 2025.
[Online]. 
Available: https://arxiv.org/abs/2507.18733




\bibitem{ref_WMMSE}
Q. Shi, M. Razaviyayn, Z.-Q. Luo, and C. He, 
``An iteratively weighted MMSE approach to distributed sum-utility maximization for a MIMO interfering broadcast channel,'' 
\emph{IEEE Trans. Signal Process.}, 
vol. 59, no. 9, pp. 4331$-$4340, Sep. 2011.



\bibitem{ref_MM}
Y. Sun, P. Babu, and D. P. Palomar,
``Majorization-minimization algorithms in signal processing, communications, and machine learning,''
\emph{IEEE Trans. Signal Process.},
vol. 65, no. 3, pp. 794$-$816, Feb. 2017.


\bibitem{ref_ADMM}
S. Boyd, N. Parikh, E. Chu, B. Peleato, and J. Eckstein, 
``Distributed optimization and statistical learning via the alternating direction method of multipliers,'' 
in \emph{Found. and Trends in Machine Learning}, 
vol. 3, no. 1, pp. 1$-$122, 2011.

\bibitem{ref_TMA}
G. Ni, C. He, Y. Gao, J. Chen, and R. Jin, 
``High-efficiency modulation and harmonic beam scanning in time-modulated array,'' 
\emph{IEEE Trans. Antennas Propag.}, 
vol. 71, no. 1, pp. 368$-$380, Jan. 2023.




\bibitem{ref_channel_estimation_1}
Z. Li \emph{et al.}, 
``Transmissive RIS transceiver enabled multistream communication systems: Design, optimization, and analysis,'' 
\emph{IEEE Internet Things J.}, 
vol. 12, no. 5, pp. 5985$-$6000,  Mar. 2025.


\bibitem{ref_channel_estimation_2}
G. Zhou, C. Pan, H. Ren, P. Popovski, and A. L. Swindlehurst, 
``Channel estimation for RIS-aided multiuser millimeter-wave systems,'' 
\emph{IEEE Trans. Signal Process.}, 
vol. 70, pp. 1478$-$1492, Mar. 2022.


\bibitem{ref_BCA}
D. P. Bertsekas,
``Nonlinear programming,''
\emph{Journal of the Operational Research Society},
vol. 48, no. 3, pp. 334$-$334, 1997.

\bibitem{ref_CVX}
M. Grant and S. Boyd,
\emph{CVX: Matlab software for disciplined convex programming}, 
version 2.1, http://cvxr.com/cvx, Mar. 2014.


\bibitem{ref_Convex Optimization}
S. Boyd and L. Vandenberghe,
\emph{Convex Optimization.}
New York: Cambridge University Press, 2004.







\end{thebibliography}
\end{document}